\documentclass[sigconf,nonacm]{acmart}

\usepackage{microtype}
\usepackage{graphicx}
\usepackage{subfigure}
\usepackage{booktabs}

\usepackage{setspace}

\usepackage{amsmath}
\usepackage{mathtools}
\usepackage{algorithmic}
\usepackage{graphicx}
\usepackage{textcomp}
\usepackage{xcolor}
\usepackage{amsfonts}
\usepackage{epsfig}
\usepackage{algorithm}
\usepackage{wrapfig}
\usepackage{balance}
\usepackage{url}
\usepackage{mathtools}
\usepackage{natbib}
\usepackage{multirow}
\usepackage{soul,comment}

\usepackage{breakcites}

\usepackage[T1]{fontenc}
\usepackage{enumerate}
\usepackage{inputenc}

\usepackage{graphicx} 
\usepackage{subfigure}

\usepackage{booktabs,balance}
\usepackage{rotating}
\usepackage{boldline}
\usepackage{makecell}
\usepackage{multirow}
\usepackage{balance}

\usepackage{tikz}

\newcommand{\bsmat}{\begin{bmatrix} }
\newcommand{\esmat}{\end{bmatrix} }

\usepackage{environ}
\NewEnviron{smallequation}{%
    \begin{equation}
    \scalebox{0.97}{$\BODY$}
    \end{equation}
    }
    \NewEnviron{smallalign}{%
    \begin{equation}
    \scalebox{0.97}{$\BODY$}
    \end{equation}
    }

\title{GUITAR: Gradient Pruning toward Fast Neural Ranking\vspace{0.1in}}

\author[Weijie Zhao, Shulong Tan, Ping Li]{{Weijie Zhao, Shulong Tan, Ping Li}\\
Cognitive Computing Lab\\
Baidu Research\\
10900 NE 8th St. Bellevue, WA 98004, USA\\
  \{zhaoweijie12, tanshulong2011,  pingli98\}@gmail.com
}

\begin{abstract}\vspace{0.1in}
\noindent With the continuous popularity of deep learning and representation learning, fast vector search becomes a vital task in various ranking/retrieval based applications, say recommendation, ads ranking and question answering.
Neural network based ranking is widely adopted due to its powerful capacity in modeling complex relationships, such as between users and items, questions and answers.
However, it is usually exploited in offline or re-ranking manners for it is time-consuming in computations.
Online neural network ranking--so called \textbf{fast neural ranking}--is considered challenging because neural network measures are usually non-convex and asymmetric. Traditional Approximate Nearest Neighbor (ANN) search which usually focuses on metric ranking measures, is not applicable to these advanced measures.

In this paper, we introduce a novel graph searching framework to accelerate the searching in the fast neural ranking problem. The proposed graph searching algorithm is bi-level: we first construct a probable candidate set; then we only evaluate the neural network measure over the probable candidate set instead of evaluating the neural network over all neighbors. Specifically, we propose a gradient-based algorithm that approximates the rank of the neural network matching score to construct the probable candidate set; and we present an angle-based heuristic procedure to adaptively identify the proper size of the probable candidate set. Empirical~results on public data confirm the effectiveness of our proposed~algorithms.
\end{abstract}
\begin{document}

\maketitle

\section{Introduction}\label{sec:intro}

With the rapid development of deep neural networks, recently, neural network based ranking models attract great attention in information retrieval, advertising, knowledge graphs,  advertising, recommendation, and question answering systems~\citep{guo2016deep,covington2016deep, xiong2017end, he2017neural, chen2017reading,devlin2019bert,fan2019mobius, chang2020pre,tan2020fast,guo2020deep,li2020video, tan2021fast,fei2021gemnn, yu2022boost,yu2022egm}.
The neural network based ranking model takes a query vector and a base vector as input and predicts a ranking score that measures how the query vector matches the base vector.
Neural networks are flexible in modeling complex relationships among different kinds of objects, such as queries and documents, users and items, or questions and answers. For example, one neural network can model how likely a user (query vector) will be interested in an item (base vector).
Given a user vector, ideally, we can enumerate all item vectors and compute their matching score with the neural network to recommend the most relevant items to the user. However, real-world recommendation systems forbid the enumeration, as it is often too time-consuming to deploy for online ranking applications over a large number of items. Therefore, neural network based ranking models are usually used in offline ranking or re-ranking on pre-produced small subsets~\citep{covington2016deep,chen2017reading,chang2020pre}.
On the other hand, in order to apply neural network matching measures in online applications, machine learning engineers design a bi-level retrieval workflow: they first retrieve a small subset of item candidates through simple similarity measures, e.g., cosine similarity, using efficient searching indices, like \textit{approximate nearest neighbor (ANN)} search methods~\citep{friedman1975algorithm, friedman1977algorithm, broder1997syntactic, indyk1998approximate,  broder1998min, gionis1999similarity,charikar2002similarity, datar2004locality, li2005using,  jegou2011product, li2012one, li2013sign, zhou2019mobius, malkov2020efficient,zhao2020song,li2022c}; then the neural network matching measures are employed over the small candidate subset to generate final ranking and recommendation.

\vspace{0.1in}
\noindent\textbf{Fast neural ranking.} Fast neural ranking is proposed to use ANN methods with neural network matching measures to perform approximate ranking by neural network measures~\citep{mitra2018introduction}. The fast neural ranking problem is considered challenging since these ranking measures are complex, usually non-convex and asymmetric. Conventional ANN search methods are designed for some simple ranking measures, such as $\ell_2$ distance, Cosine similarity, or inner product. It is not straightforward to extend ANN search methods for fast neural ranking scenarios.

\citet{tan2020fast} show that building a proximity graph over item vectors using $\ell_2$ distance and performing a graph-based ANN search on the graph results a coordinate descent on the item space for the neural network measure. Specifically, \citet{tan2020fast} extend the definition of traditional greedy search to a generic setting:
Let $X$ and $Y$ be subsets of Euclidean spaces, given a data set $S = \{x_1, \dots, x_n\} \subset X$ and a matching function (which can be a neural network), $f: X\times Y \to \mathbb R$, given $q\in Y$, we target to find:
    \begin{align} \label{eq:obfs}
    \arg\max_{x_i\in S} f(x_i, q), \quad \text{ for } q\in Y.
    \end{align}

There are no strong assumptions for the matching function: metric or non-metric, linear or non-linear, convex or non-convex, symmetric or asymmetric. The distance measures focused by ANN search or any learned neural network based measures are all special cases of the matching function. Beyond the definition, they also provide a solution for fast neural ranking (the case when the matching function is a neural netowrk), called Search on L2 Graph (SL2G). They extend traditional graph based fast vector searching algorithms by constructing graph indices in $\ell_2$ distance but searching according to the neural network measure.

\begin{figure}[htbp]
\centering
    \includegraphics[width=3in]{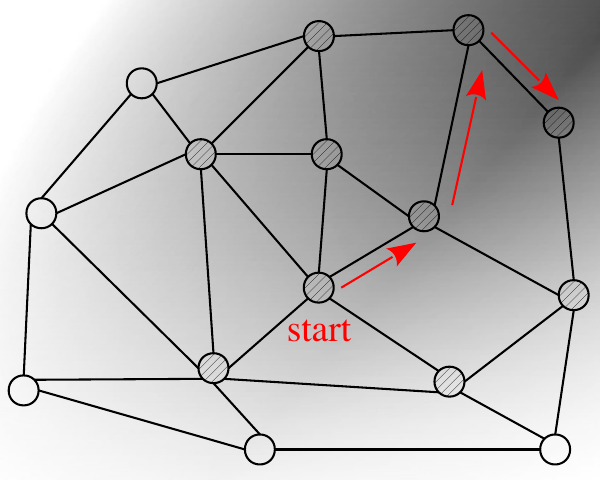}
    \caption{An example for the fast neural ranking searching procedure. Each vertex represents a base vector, e.g., an item embedding vector. For a given query vector $q$, e.g., a user vector, we have a neural network $f(x,q)$ to evaluate the similarity or ranking score of a base vector $x$ and $q$. The value of $f(x,q)$ is color encoded over the space of $x$ (since $q$ is fixed for a given query): the darker the color is, the higher the similarity of $x$ and $q$ we have. The searching begins from the start point, computes $f(x,q)$ for each neighbor, and makes the neighbor with the highest similarity be the next frontier of the searching. The searching is performed iteratively until no $x$ with a higher similarity can be found. The hatched vertices represent the vectors that we computed neural network measure $f(x,q)$ with. We can observe that along the searching path, many $f(x,q)$ computations are wasted since they are far away from the desired searching direction.}\label{fig:search-example}
\end{figure}

Figure~\ref{fig:search-example} presents an example of the SL2G searching algorithm. The searching procedure follows the A$^{*}$ heuristic fashion as most graph-based ANN algorithms: in each iteration, we extract a frontier vertex from a priority queue, probe all neighbors of the current frontier vertex, and insert those neighbor vertex into the priority queue. The weight of each vertex in the priority queue is its similarity to the query vertex.

This searching algorithm works well for common similarity or distance measures.
However, the evaluation of a neural network is substantially more complicated and time-consuming than a common measure, e.g., $\ell_2$---evaluating a neural network needs to compute multiple matrix multiplications while computing a common measure only requires to go over all dimensions of two vectors once and calculate an aggregation.
As a result, in the retrieval problem under neural network similarity measures, the number of neural network evaluations becomes a dominant factor in the execution.

\vspace{0.1in}
\noindent\textbf{Intuition.} One question can be straightforwardly raised: \textit{Can we reduce the number of neural network evaluations to improve the search performance?} Recall the visual illustration in Figure~\ref{fig:search-example}, we mark the vertices that are evaluated with the neural network measures as hatched. The actual searching path (red arrows) usually follows the vertex that has the highest matching score.
Meanwhile, a large proportion of the evaluated vertices are not used in the later search iterations.
This observation provides us with an opportunity to evaluate a fewer number of neural network measures: we can first find a \textit{probable candidate set} which is a small subset of neighbors that contains the neighbor with the highest score, then we only need to evaluate the neural network on this subset of vectors to determine the searching frontiers of the following iterations.

\vspace{0.1in}
\noindent\textbf{Challenges \& approaches.}
We identify two major challenges in locating the probable candidate set. The first challenge is to propose a measure that effectively ranks the neighbors without evaluating the neural network measure for all neighbors. The rank should approximate the order of the neural network measure so that we can obtain the neighbor vertex with the highest matching score from the forepart of the ranked neighbors.
We introduce a gradient-based approximation to rank the neighbor vertices that only requires us to compute the gradient once and rank the neighbor vertices using their spatial relationship to the gradient, e.g., separation angle.
The second challenge is to identify the size of the probable candidate set. When the subset is too small, the required neighbor vertex with the best matching score may not reside in the subset. On the other hand, when the subset is too large, although the best candidate is very likely to be included in the subset, we have to evaluate the neural network measure on all vertices in this subset---it is still time-consuming. We propose an angle-based heuristic to adaptively select the neighbor vertices within a given angle range into the probable candidate set.

\vspace{0.05in}
\noindent\textbf{Contributions.} We summarize our contributions as follows:

\vspace{-0.01in}

\begin{itemize}
    \item We introduce \texttt{GUITAR} (acronym for Gradient prUnIng Toward fAst Retrieval), a novel graph searching framework to accelerate the searching in the fast neural ranking problem. The proposed graph searching algorithm is bi-level: we first construct a probable candidate set; then we only evaluate the neural network measure over the probable candidate set instead of evaluating the neural network over all neighbors.\vspace{0.05in}
    \item We propose a gradient-based algorithm that approximates the rank of the neural network matching score to construct the probable candidate set.\vspace{0.05in}
    \item We present an angle-based heuristic procedure to adaptively identify the proper size of the probable candidate set.\vspace{0.05in}
    \item We extensively study our method on public datasets. The experimental results confirm the effectiveness of our solutions.
\end{itemize}


\section{Related Work}\label{sec:relwork}
In this section, we will review the background of this work. Firstly, we would like to introduce the fast neural ranking problem, which is generalized from the traditional Approximate Nearest Neighbor (ANN) search. And then, possible solutions and limitations will be discussed which motivate the proposed method.

\subsection{Fast Neural Ranking}

Fast ranking or searching is the core problem of Information Retrieval, such as top-$K$ recommender systems for e-commerce and link prediction for social networks. The queries (e.g., users in recommender systems) often have some context, say locations and time, which are unknown beforehand. So the search process is required to be conducted as an online manner. For online services, the search efficiency is as important as search effectiveness. We usually leverage some pre-computed data structures or intermediate data formats (or called indexing) to speed up the online ranking/searching, in sub-linear complexity. Inverted index is commonly applied in modern search engines, say Google search and Bing search. Inverted indices are generally designed for discrete text data. However, there is more and more dense vector data, with the popularization of deep learning and representation learning. Inverted indices are not good at indexing these kinds of dense vector data. Therefore, a series of \textbf{fast vector search} methods are proposed and widely applied, such as hashing based methods~\citep{indyk1998approximate,gionis1999similarity,charikar2002similarity, rajaraman2011mining,shrivastava2012fast,li2012one,li2013sign,li2022c}, quantization based methods~\citep{jegou2008hamming, jegou2011product, ge2013optimized, wu2017multiscale}, graph based indices~\citep{hajebi2011fast,wu2014fast,iwasaki2016pruned,iwasaki2018optimization,zhou2019mobius,malkov2020efficient,tan2021norm} and tree based methods~\citep{friedman1975algorithm,friedman1977algorithm, cayton2008fast,curtin2013fast,curtin2014dual}.

Traditional fast vector search methods focus on ranking by metric measures (e.g., cosine similarity and $\ell_2$ distance). Searching beyond metric measures was considered challenging.
Only a few simple non-metric measures were studied in previous literature. Among which searching by the inner product is popular because of its wide applicability in recommendation and classification tasks. Quite a few algorithms are proposed for inner product search~\citep{ram2012maximum,bachrach2014speeding,shrivastava2014asymmetric,shrivastava2015asymmetric,yu2017greedy,yan2018norm,morozov2018non,zhou2019mobius,tan2021norm}, which is usually referred as Maximum Inner Product Search (MIPS) in the literature.  The Mercer kernels are the extension of inner product in Hilbert space $\mathcal{H}$. Efficient searching methods by Mercer kernels are also proposed~\citep{curtin2013fast,curtin2014dual}.

Recently, more and more information retrieval systems exploit neural network based ranking models~\citep{guo2016deep, lu2013deep,dehghani2017neural,chang2020pre,tan2020fast,tan2021fast,yu2022egm} to take advantage of their powerful capacities in modeling complex relationships between objects, say semantic information between searching queries and documents. These prediction models usually produce neural network based ranking measures. Setting a user vector and an item vector as inputs, one can design any neural network structures as ranking measures. Parameters of the network are learned on training data but not fixed beforehand. These kinds of neural network based searching functions are usually non-convex, which are not studied by traditional ANN search work. These are lots of real cases of neural networks based searching measures, such as Multi-Layer Perceptron (MLP) and BERT-style ones~\citep{he2017neural,tay2018latent,severyn2015learning,dehghani2017neural}, which has wide applications in recommendation, advertising (ads) ranking, knowledge graphs, and retrieve based question answering~\citep{guo2016deep,   xiong2017end, guo2020deep, covington2016deep, he2017neural, chen2017reading,devlin2019bert,fan2019mobius, li2020video, chang2020pre,fei2021gemnn,yu2022boost,yu2022egm}. Efficient or sub-linear ranking by neural network based measures, or called \textbf{fast neural ranking}, has a very broad application value. But it is still an open research question due to the  essential complexity.

\subsection{Previous  Methods for Neural Ranking}

Traditional ANN search algorithms typically do not extend to complex similarity measures such as neutral ranking. For example, with hashing based ANN algorithms, usually one particular  method is  designed for one specific metric measure: the celebrated  ``sign random projection'' (SgnRP) is designed for cosine similarity~\citep{goemans1995improved,charikar2002similarity}, the ``sign Cauchy projections'' is for the chi-square similarity~\citep{li2013sign}, and the ``minwise hashing'' is for Jaccard similarity~\citep{broder1997syntactic,broder1998min,charikar2002similarity,shrivastava2012fast,li2012one,li2022c}. Ball tree-based methods are proposed for searching by some non-metric measures, such as Max-kernel search~\citep{curtin2013fast,curtin2014dual} and searching by Bregman divergence~\citep{cayton2008fast}. But it is infeasible to be extended for neural network based measures.

Search on graph methods often claim that there are no constraints on searching measures (actually must be symmetric)~\citep{hajebi2011fast,morozov2018non,zhou2019mobius, malkov2020efficient,zhao2020song, tan2021norm}. It was shown by~\cite{tan2020fast,tan2021fast} that the performance is quite limited when we apply traditional  graph based methods in fast neural ranking. On the other hand, the \textbf{Search on L2 Graph (SL2G)} method~\citep{tan2020fast} is designed for the fast neural ranking problem. 

\vspace{0.05in}
The basic idea of SL2G is the following:
\begin{enumerate}
    \item[1)] No matter what the given binary function $f$ is, they construct a Delaunay graph (or an approximate one) with respect to  $\ell_2$ distance (which is defined on searching/base data $X$ and independent of queries) in the indexing step.\vspace{0.05in}
     \item[2)] Then SL2G performs the greedy search on this graph by the binary function $f$ in the searching step.
	\end{enumerate}

\vspace{0.05in}

The theoretical basis of SL2G is that, the performance of greedy search on $\ell_2$ graph is similar to optimizing fast neural ranking by ``coordinate'' descent in Euclidean space. If the $f$ is smooth and the data are dense enough, SL2G will reach an approximate local optimum. The indexing construction (i.e., via $\ell_2$ distance) is very efficient and independent from the searching measure $f$. However, it requires to compute the function $f$ multiple times at each searching step (Details can be found in Figure~\ref{fig:search-example}). So SL2G is still not applicable to very complicated ranking measures on large data. BEGIN~\citep{tan2021fast} includes the binary function $f$ (the neural network) into the index construction by building a bipartite graph. BEGIN trades a more time-consuming offline indexing for better constructed graph for online querying. Nonetheless, BEGIN also requires us to evaluate function $f$ for all neighboring vertices during searching.

In this paper, we propose a searching algorithm (GUITAR) that reduces the number of neural network evaluations to accelerate the searching performance. With the proposed fast searching algorithm, lots of generic searching measures can be applied for online ranking services, say those neural network based measures~\citep{he2017neural,tay2018latent,severyn2015learning,dehghani2017neural,xiong2017end}. More advanced semantic information will be captured in the ranking/searching procedure and
the user experience would be improved greatly.
Besides, some common optimizations can be added onto the proposed method to further improve the searching efficiency: (1) take advantage of GPU to accelerate the computations further~\citep{johnson2021billion,zhao2020song}; (2) compress the neural network based measures by knowledge distillation techniques~\citep{hinton2015distilling,tang2018ranking,zhang2018deep}.

\vspace{0.05in}

\section{GUITAR}\label{sec:method}
We present our proposed gradient walk searching algorithm in this section. We begin with an overview of the GUITAR searching framework. Then we introduce the details of each stage in the framework step by step.

\subsection{Overview}\label{sec:overview}
Our proposed GUITAR searching framework introduces a bi-level filtering into the conventional graph-based Greedy searching algorithm~\citep{malkov2020efficient}. In each searching iteration, the conventional searching algorithm requires us to compute the distance/similarity for all neighbors of the current searching frontier. The searching becomes slow when the distance/similarity is time-consuming to compute. Our proposed bi-level searching filters out the neighbor vertices that are ``unlikely'' to be on the searching path to the top-k results and only evaluates the time-consuming neural network measure on the probable candidate set.

\newpage

\begin{algorithm}[htbp]
\caption{Bi-Level Graph Searching}\label{alg:search}
\begin{flushleft}
\textbf{Input:} query vector $q$; neural network measure $f(x,q)$; search starting point $start$.
\end{flushleft}
\begin{algorithmic}[1]
\STATE Init a priority queue $\textit{pq}$ with $\textit{start}$ whose priority is $f(\textit{start,q})$
\STATE $\textit{visited} \leftarrow \{\textit{start}\}$
\WHILE{$\textit{pq} \not= \emptyset$}
    \STATE $\textit{frontier} \leftarrow \textit{extract\_top}(\textit{pq})$
    \IF{$f(\textit{frontier})$ is worse than any element in top-k}
        \STATE \textbf{break}
    \ENDIF
    \STATE update top-k result with $\textit{frontier}$
    \STATE $\textit{ranked} \leftarrow \textit{rank\_neighbors}(\textit{frontier})$ \label{line:rank}
    \STATE $\textit{probable} \leftarrow \textit{prune\_neighbors}(\textit{ranked})$ \label{line:prune}
    \FOR{\textbf{each} vertex $v$ in \textit{probable}}
        \IF{$v \not\in \textit{visited}$}
            \STATE insert $v$ into $pq$ with priority $f(v,q)$
            \STATE $\textit{visited} \leftarrow \textit{visited} \cup \{v\}$
        \ENDIF
    \ENDFOR
\ENDWHILE
\RETURN found top-k result
\end{algorithmic}
\end{algorithm}

We present the workflow of the bi-level graph searching in Algorithm~\ref{alg:search}. The major searching structure follows the convention graph-based ANN searching algorithm except two places: \textit{rank\_neighbors} in Line~\ref{line:rank} and \textit{prune\_neighbors} in Line~\ref{line:prune}. The \textit{rank\_neighbors} procedure targets to rank the neighbor vertices, such that the vertices with higher ranks include the neighbor vertices having the highest matching scores under the neural network measure. The major challenge here is to have a ranking algorithm that efficiently approximates the neural network rank without actually evaluating the neural network measure. We illustrate the details of the ranking algorithm in Section~\ref{sec:rank}. On the other hand, the \textit{prune\_neighbors} procedure determines the number of vertices that we will evaluate neural network measure on. When the number is too small---only the very highly ranked vertices are included in the search, we may not cover the required neighbor vertex with the best matching score. Also, when the number is too large, we have to evaluate the neural network measure on the large number of vertices. We discuss this trade-off and introduce our selection criterion in Section~\ref{sec:prune}.

\subsection{Ranking Neighbor Vertices}\label{sec:rank}
In this section, we present our neighbor vertices ranking algorithm for the \textit{rank\_neighbors} procedure in Algorithm~\ref{alg:search}. Our goal is to rank the neighbor vertices of the current searching frontier such that the highest ranked vertices include the neighbor vertices with the highest matching score under the neural network measure. The core question we have to answer is: \textit{How to approximate the neural network measure without actually evaluating it?}

\vspace{0.1in}
\noindent\textbf{Gradient.}
By the definition of gradient, given a loss function $L$ and a point $p$, the gradient represents the direction and rate of fastest increase at point $p$. When the gradient is non-zero, the direction of the gradient represents the direction in which $L$ increases most quickly from $p$. Meanwhile, the magnitude of the gradient represents the rate of increase in that direction.
We denote the neural network measure is $f(x,q)$, where $x$ is a base vector and $q$ is the query vector---given a query vector, the fast neural ranking problem targets to find the best $k$ base vectors that maximize the matching score $f(x,q)$ from all base vectors.
Without loss of the generality, we assume the value range of $f(x,q)$ is $[0,1]$, i.e., a sigmoid function is applied on the output logit to map the real value to $[0,1]$, where $0$ means the lowest matching score and $1$ is the highest matching score. We define the loss function $L$ as:
\begin{equation}
L(x,q) = 1 - f(x,q).
\label{eq:loss}
\end{equation}
Here $L(x,q) \geq 0$ because $f(x,q) \leq 1$. No absolute value is required. This loss definition is equivalent to an $\ell_1$ loss.
According to the definition of gradient, the reverse direction of the gradient is the direction in which $L(x,q)$ decreases most quickly from $x$. In the context of the fast neural ranking, the parameters of the neural network $f$ and the query vector $q$ are given as input---they are fixed constants. Thus, the only parameter we can optimize to minimize $L(x,q)$ is $x$. The reverse direction of the partial derivative of $x$, $-\frac{\partial L}{\partial x}$, represents the best direction to search in next iterations.

\begin{figure}[h]
\centering
    \includegraphics[width=3in]{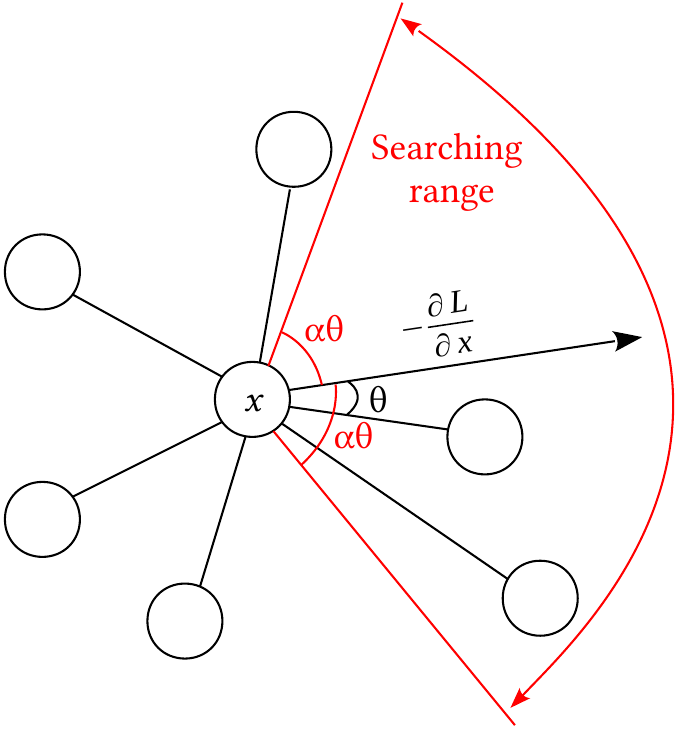}
    \caption{An example for ranking neighbor vertices and adaptive pruning on the separation angles.}\label{fig:example}
\end{figure}

\vspace{0.1in}
\noindent\textbf{Separation Angle.} This provides us with an opportunity to approximate the ranking of the neighbors: we can use the separation angle between neighbor vertex and $-\frac{\partial L}{\partial x}$ as the ranking criterion: the closer to the direction of $-\frac{\partial L}{\partial x}$, the higher the rank it has. Theoretically, when the distances between $x$ and its neighbors are sufficiently small, the neighbor $x'$ with a smaller separation angle to the direction of $-\frac{\partial L}{\partial x}$ should have a smaller $L(x',q)$, thus leading a greater matching score $f(x',q)$. In the proximity graph scenario, the neighbor vertices of $x$ are nearest neighbors of $x$. Empirically, we can assume that they are sufficiently close for real-world datasets and applications. Figure~\ref{fig:example} depicts a visual illustration of the separation angle---for each neighbor vertex, we rank the vertex via its separation angle to the reverse direction of the gradient (i.e., $-\frac{\partial L}{\partial x}$).

\vspace{0.1in}
\noindent\textbf{Projection.} In the gradient descent context, the step size (learning rate) determines how far a step is performed on the reverse direction of the gradient in one iteration. Similarly in our ranking problem, for each neighbor vertex $x'$, we may also take its distance to $x$ into consideration, as well as its separation angle to $-\frac{\partial L}{\partial x}$: We can use its projection to $-\frac{\partial L}{\partial x}$ as an alternative to the separation angle based ranking. Since $x$ and $x'$ are neighbors in the proximity graph, $x$ is likely to share a similar distance to each neighbor vertex $x'$. Therefore, we assume the ranking difference between separation angle and projection is minimal. We investigate the effect of both ranking criteria in Section~\ref{sec:exp}.

\vspace{0.1in}
\noindent\textbf{Choices of loss function.} Note that the choice of different norms in loss functions $L$ does not have any impact on the direction of the gradient. For example, if we change the loss function with a $\ell_2$ norm (instead of $\ell_1$ we used in Equation~\eqref{eq:loss}), the only difference in the gradient is that a scalar factor $2f(x,q)$ is attached to the magnitude of the gradient. Likewise, we can use $L(x,q)=-f(x,q)$ as the loss function for non-capped matching scores to obtain exactly the same gradient direction as our loss function in Equation~\eqref{eq:loss}.

\vspace{0.1in}
\noindent\textbf{Cost of the ranking.} Our gradient based ranking method does not require us to evaluate the neural network measures on all neighbor vertices. The primary cost of the \textit{rank\_neighbors} comes from three parts: (a) gradient computation; (b) separation angle computation; and (c) sorting. We denote the neural network evaluation time is $F$, the average number of neighbors of a vertex is $B$, and the base vector dimension is $D$. The gradient computation takes around $2F$ time---one feed forward and one backward to compute the gradient for $x$. After that, we need to compute the separation angles for all neighbor vertices---it takes $O(BD)$ time. Finally, we sort the $B$ separation angles in $O(B\log{B})$ time. Overall, the cost of the ranking is $2F + O(BD + B\log{B})$. Note that $B$ is usually a small constant for proximity graphs, e.g., $32$ while the evaluation of a neural network is substantially larger than $B$, i.e., $F \gg B$. In addition, the neural network takes the $D$-dimensional vector into input and performs complicated matrix multiplications to produce the final matching score. Its cost is also substantially greater than $D$, i.e., $F \gg D$. Therefore, the dominant factor in our ranking algorithm only comes from the evaluation of the neural network measure: $2F$. As long as we can prune out more than $2$ neural network evaluations in the following \textit{prune\_neighbors} procedure, we can benefit from the bi-level graph searching framework.

\vspace{0.1in}

\subsection{Neighbor Pruning}\label{sec:prune}
\vspace{0.05in}

Now we have already efficiently ranked the neighbor vertices to approximate their ranks under the neural network measure. In order to construct the probable candidate set, we still have one challenge left: \textit{How to properly choose the size of the probable candidate set?}

\vspace{0.1in}
\noindent\textbf{Fixed constant.} The most straightforward solution is to select a constant number or a constant proportion of the highest ranked neighbor vertices to form the probable candidate set. The constant pruning solutions ignore the separation angle distribution: when a lot of neighbor vertices are close to the direction of $-\frac{\partial L}{\partial x}$, the constant pruning strategy discards the neighbor vertices that are potentially good candidates; when only a limited number of neighbor vertices are close to the direction of $-\frac{\partial L}{\partial x}$, the constant cut-off threshold also includes unnecessary neighbor vertices that diverges too much from the desired direction. Therefore, the probable candidate set should be adaptively selected to achieve the optimal searching performance.

\vspace{0.1in}
\noindent\textbf{Adaptive search range selection.} We propose a heuristic pruning strategy that takes the separation angle of the closest neighbor vertex into consideration. Figure~\ref{fig:example} depicts a visual illustration for the strategy. As shown in the figure, we denote the closest separation angle to the direction of $-\frac{\partial L}{\partial x}$ as $\theta$. The searching range is $\alpha\theta$ that spans around the direction of $-\frac{\partial L}{\partial x}$ as $\theta$, where $\alpha$ ($\alpha \geq 1$) is a tunable tolerance parameter that represents how tolerant we are for the neighbor vertices which are not the closest.
We select the neighbor vertices within the searching range as the probable candidate set. The greater $\alpha$ is, the more vertices are selected into the set. Formally, given a query $q$ and the current searching frontier vertex $x$, the probable candidate set selection can be written as:

\vspace{0.05in}

\begin{equation}
    \textit{probable\_candidate\_set} = \left\{ x' \left\vert \text{arccos}\left(\frac{-\frac{\partial L(x,q)}{\partial x} \cdot (x' - x)}{ \left\vert -\frac{\partial L(x,q)}{\partial x} \right\vert \cdot \left\vert x' - x\right\vert} \right) \leq \alpha \theta \right.\right\},
\label{eq:angle-prune}
\end{equation}\vspace{0.05in}

\noindent where $x'$ belongs to the neighbor vertices of $x$ and $x'-x$ represents the vector starting from $x$ and ending at $x'$.

This proposed adaptive search range selection strategy tackles the disadvantages incurred in the previously discussed fixed constant pruning: only the neighbor vertices are close to the best found neighbor vertices are selected for the further neural network measure evaluation.

\vspace{0.1in}
\noindent\textbf{Adaptive selection for projection ranking.} As we discussed in Section~\ref{sec:rank}, for a neighbor vertex, we can use its projection value on the reverse direction of the gradient as an alternative ranking criterion. Similar to the adaptive separation angle selection strategy in Equation~\eqref{eq:angle-prune}, we have:

\vspace{0.05in}

\begin{equation}
    \textit{probable\_candidate\_set} = \left\{ x' \left\vert \frac{-\frac{\partial L(x,q)}{\partial x} \cdot (x' - x)}{ \left\vert -\frac{\partial L(x,q)}{\partial x} \right\vert} \geq \frac{\theta}{\alpha} \right.\right\}.
\label{eq:proj-prune}
\end{equation}\vspace{0.05in}

\noindent In the projection case, the best candidate among neighbor vertices has the greatest projection value (instead of having the smallest separation degree in the adaptive degree selection strategy). Thus, the inequality sign in Equation~\eqref{eq:proj-prune} is reversed to ``$\geq$'' and the tolerance factor $\alpha$ is put as a denominator. We empirically evaluate both ranking and pruning strategies in Section~\ref{sec:exp}.

\begin{figure*}[t]
\centering
\includegraphics[width=7in]{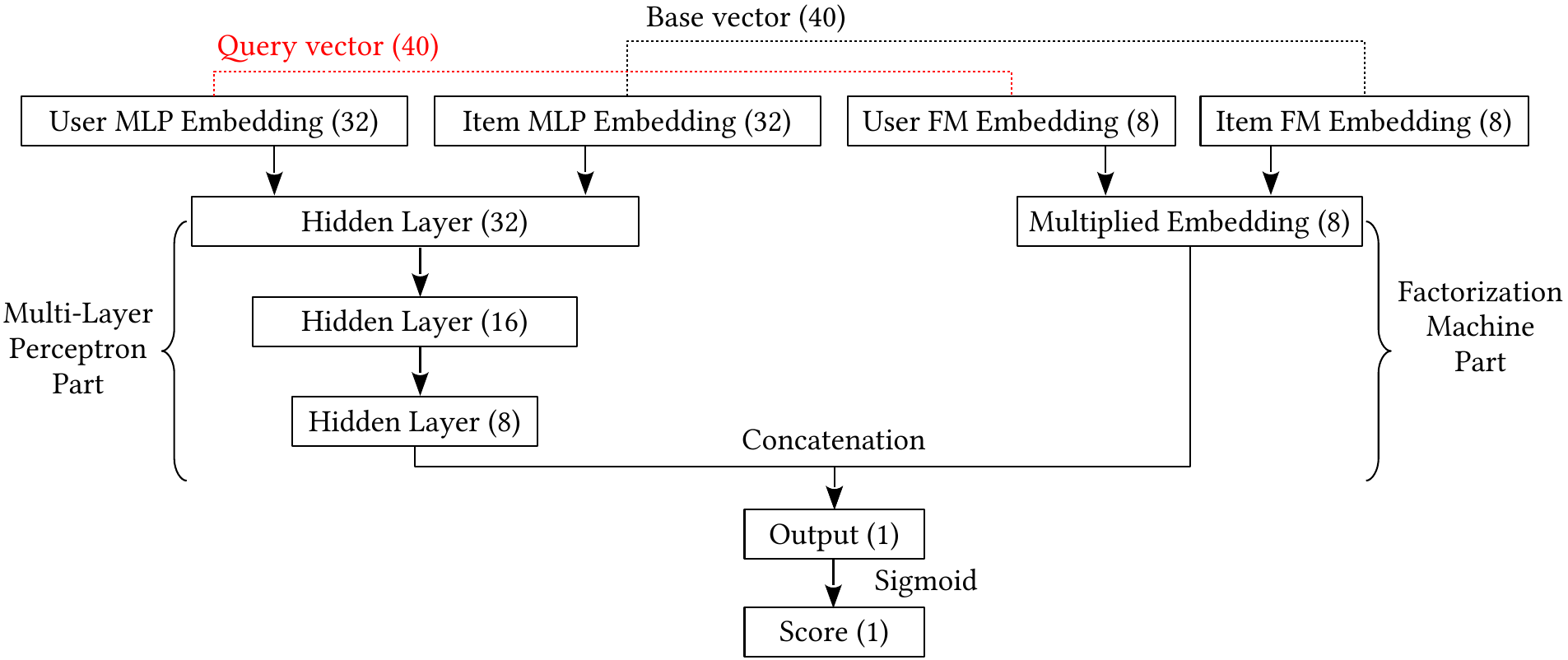}

\caption{DeepFM neural network architecture.}
\label{fig:deepfm}\vspace{0.05in}
\end{figure*}

\newpage

\section{Experiments}\label{sec:exp}
In this section, we experimentally evaluate the effectiveness of our proposed GUITAR with public recommendation system datasets under real-world neural network measures. Specifically, the experiments are targeted to answer the following questions:
\begin{itemize}
\item How is the overall performance of GUITAR compared with the state-of-the-art fast neural ranking algorithm, i.e., SL2G?\vspace{0.05in}
\item How should we set the tolerance factor $\alpha$?\vspace{0.05in}
\item How many neural network measure evaluation is saved by using GUITAR? Does it cover the overhead to utilize the bi-level searching algorithm?\vspace{0.05in}
\item Which pruning strategy is better, separation angle or projection?\vspace{0.05in}
\item Does GUITAR work on other graph indices, i.e., BEGIN?
\end{itemize}


\begin{figure*}[t]
\begin{center}
\mbox{\hspace{-0.05in}
    \includegraphics[scale = 0.32]{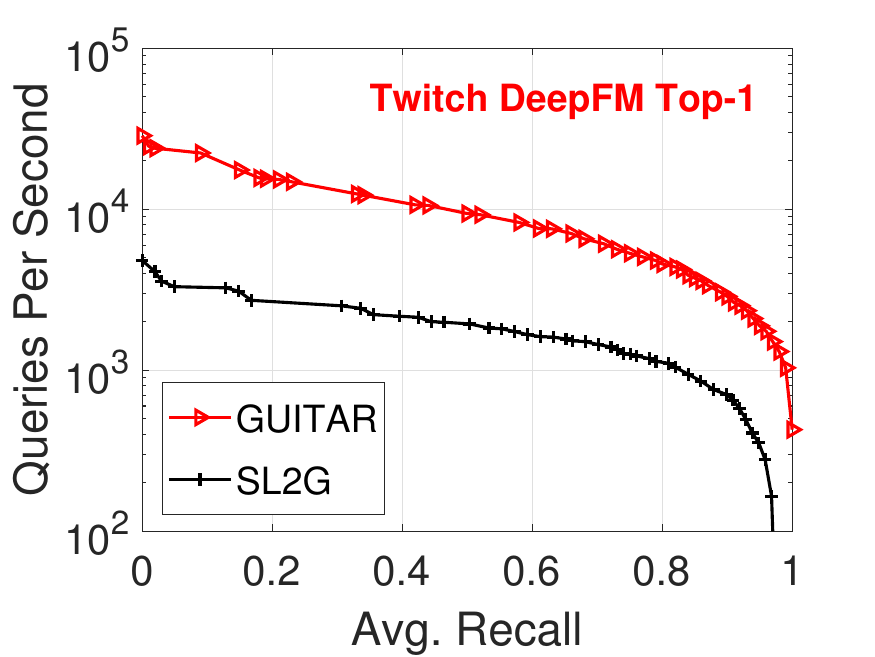}
    \hspace{-0.15in}
    \includegraphics[scale = 0.32]{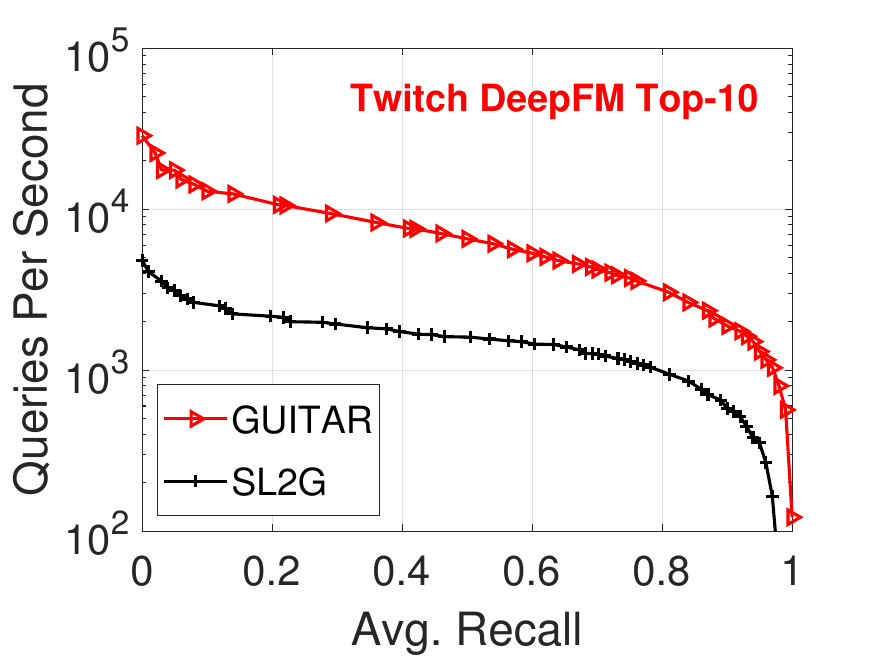}
    \hspace{-0.15in}
    \includegraphics[scale = 0.32]{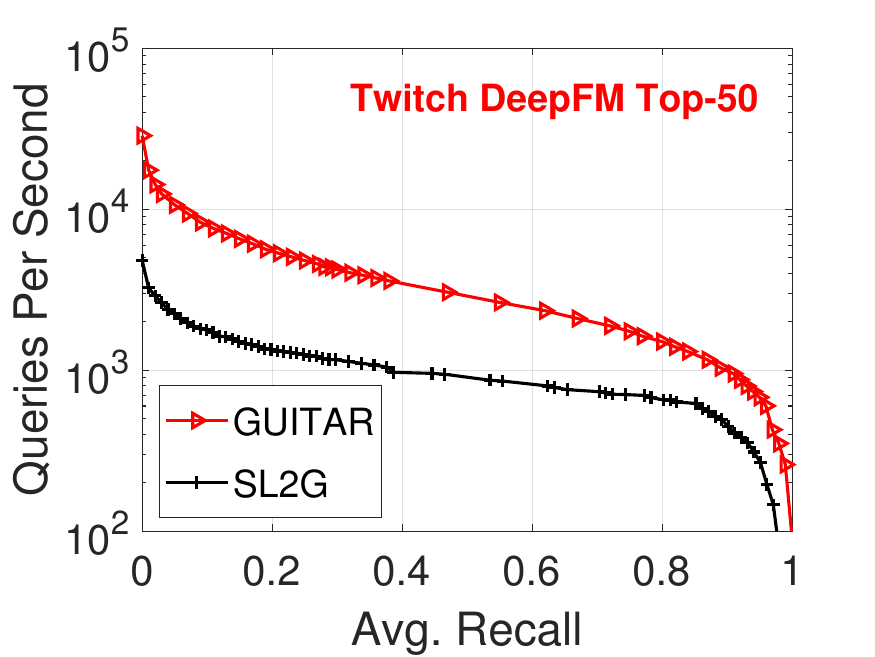}
    \hspace{-0.15in}
    \includegraphics[scale = 0.32]{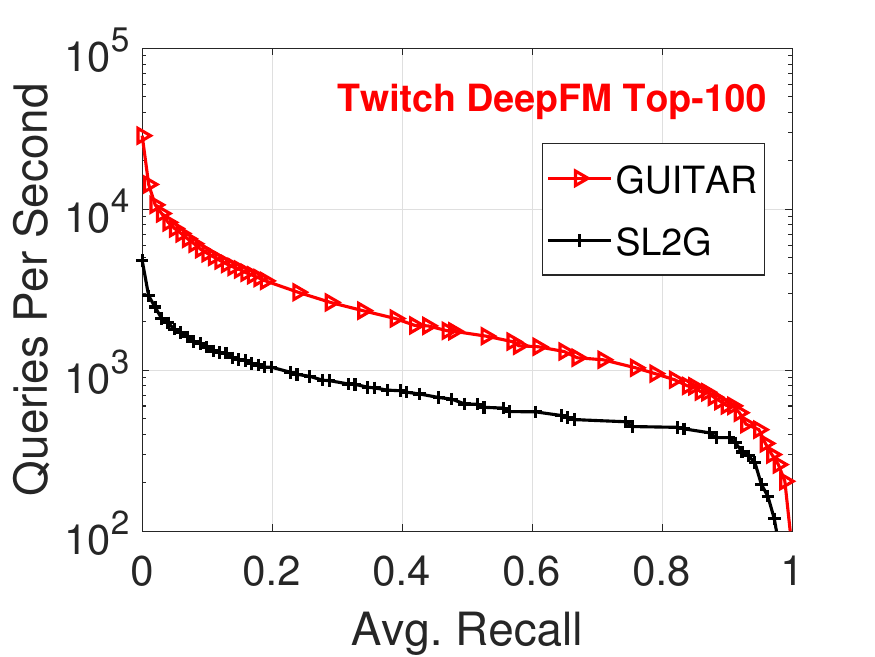}
    }
\mbox{\hspace{-0.05in}
    \includegraphics[scale = 0.32]{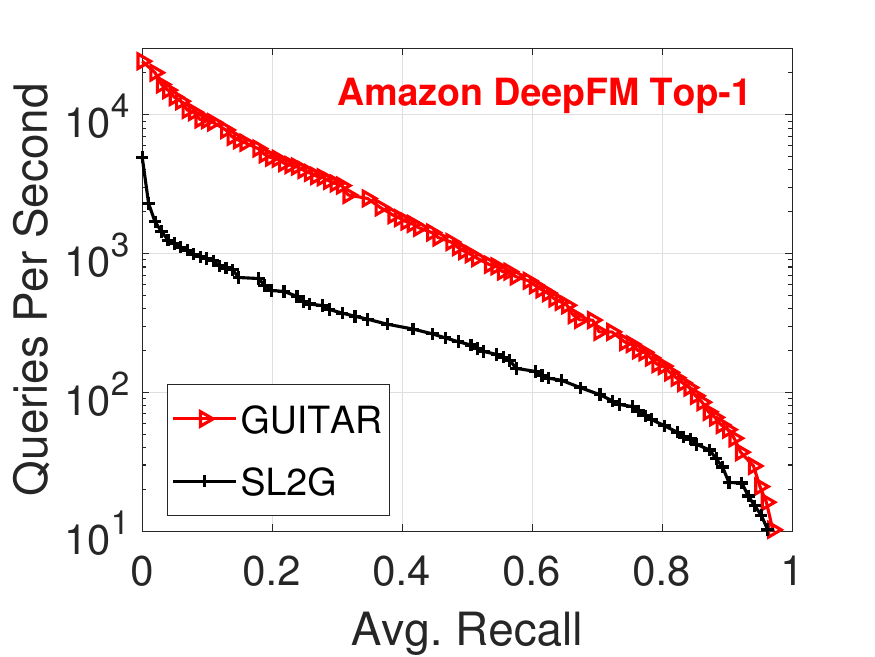}
    \hspace{-0.15in}
    \includegraphics[scale = 0.32]{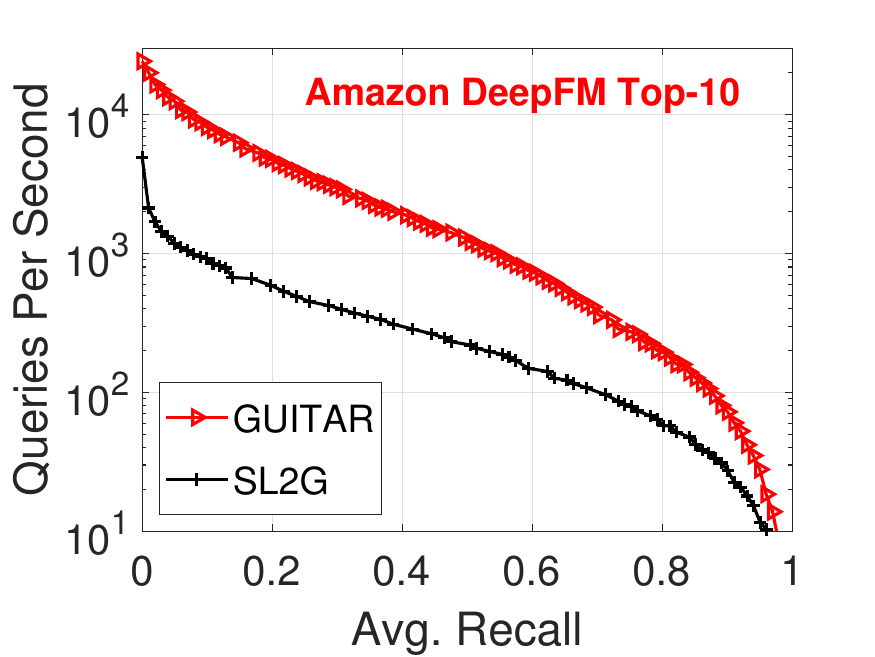}
    \hspace{-0.15in}
    \includegraphics[scale = 0.32]{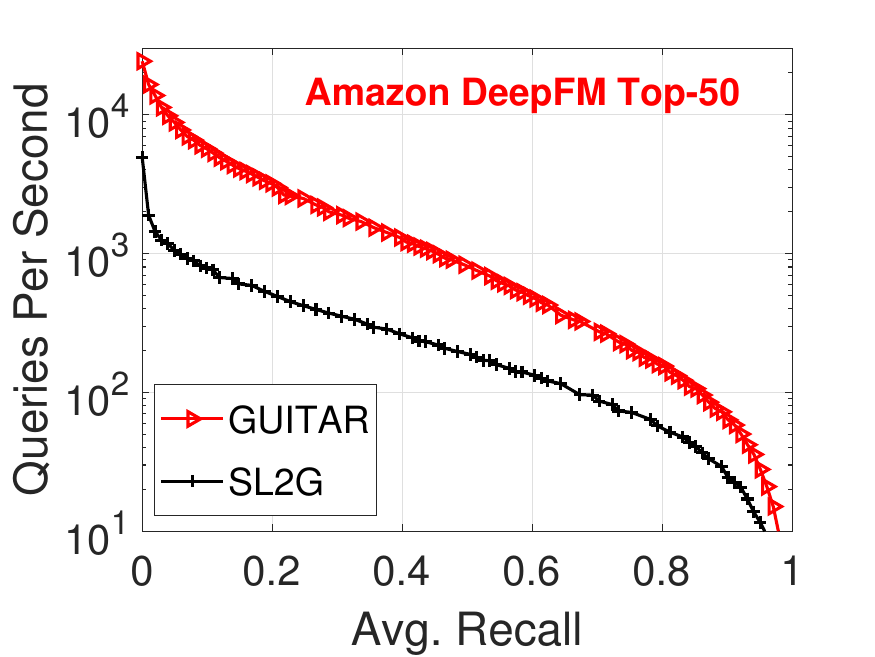}
    \hspace{-0.15in}
    \includegraphics[scale = 0.32]{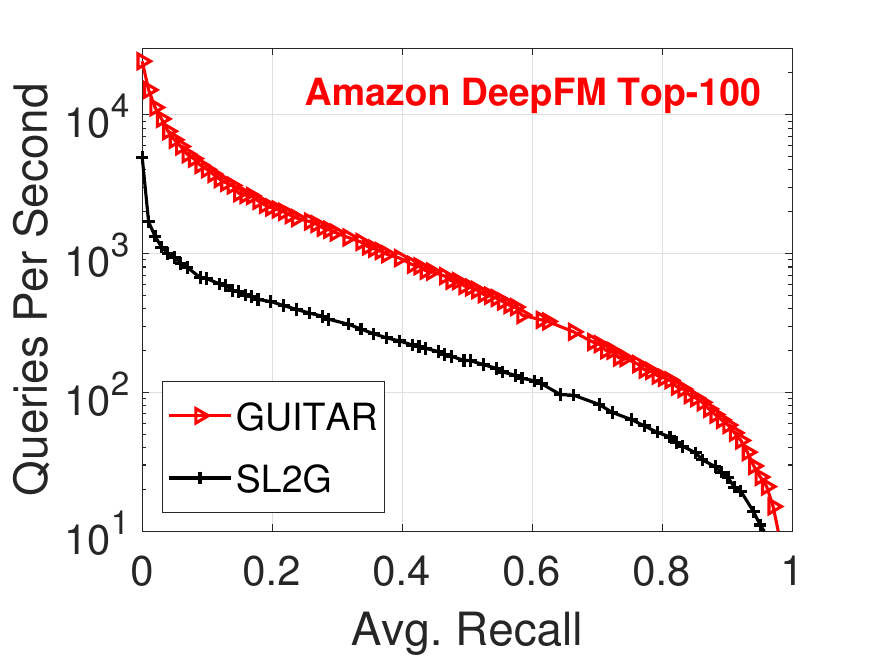}
    }
\end{center}
\caption{Recall vs. Time for top-1, 10, 50, 100 recalls. The best results are in the upper right corner. }
\label{fig:exp:guitar-sl2g}
\end{figure*}

\vspace{0.1in}
\noindent\textbf{Neural network measures.} The ranking measures used in the experiments are the DeepFM model~\citep{guo2017deepfm} trained on real world recommendation datasets. DeepFM was proposed in \citep{guo2017deepfm}, which combines the power of factorization machines for recommendation and deep learning for
feature learning in a new neural network architecture. We set the factorization part dimension as 8 and deep learning part as 32. The total dimension for users and items is both 40. Figure~\ref{fig:deepfm} depicts the detailed neural network architecture.

\vspace{0.1in}
\noindent\textbf{Datasets}. For the recommendation system datasets, we choose two with larger item sets, which are more meaningful for fast vector search: (i) \textbf{Twitch}\footnote{ \url{https://cseweb.ucsd.edu/~jmcauley/datasets.html\#twitch}}~\citep{rappaz2021recommendation} that contains users consuming streaming content on Twitch. Given a user, we target to retrieve relevant shows of steamers. (ii) Amazon Movies and TV (\textbf{Amazon})\footnote{ \url{http://jmcauley.ucsd.edu/data/amazon}} that includes the users' reviews on movies and TV shows on Amazon. For this dataset, given a user, our goal is to recommend relevant movies and TV shows to the user. To generate evaluating labels, we calculate most relevant base data points for each query by a trained neural network measure $f$.
Experiments on top-1, 10, 50 and 100 labels are recorded. After the model training, 1000 user vectors are set as testing queries. The specifications of both datasets are summarized in Table~\ref{data}.

\begin{table}[h]
\centering
\caption{Dataset Statistics.}
\begin{tabular}{|l|c|c|c|} \hline
Datasets&\# Index Vec&\# Queries&Vector Dim\\ \hline
Twitch & 739,991 &100,000 &40 \\ \hline
Amazon & 3,826,085&182,032 &40\\ \hline\end{tabular} \label{data}
\end{table}

\vspace{0.1in}
\noindent\textbf{Baseline}. We use SL2G and BEGIN as our baselines. SL2G is the state-of-the-art method that tries to solve the generic ranking problem while BEGIN focuses on the case when the neural network matching is asymmetric, e.g., user-item. Note that we do not compare with other traditional ANN search or MIPS algorithms (e.g., ANNOY
 \footnote{\url{https://github.com/spotify/annoy}}
and HNSW~\citep{malkov2020efficient}) since most of them are not designed for the fast neural ranking problem. It was demonstrated that these methods are dramatically worse than SL2G when applying them on fast neural ranking~\citep{tan2020fast}.

\vspace{0.1in}
\noindent\textbf{Implementation.} We implement GUITAR as a \texttt{C++} prototype under the code base derived from SL2G. GUITAR and SL2G share the same graph index construction algorithm and in-memory graph storage strategy. The major difference is the searching algorithm: GUITAR uses a bi-level searching algorithm to evaluate fewer times of neural network measures. The gradient computation implementation is derived from the neural network evaluation---a backward operation is implemented to obtain the gradient.

\vspace{0.1in}
\noindent\textbf{System hardware.} We execute all methods on the same computing node. The node has 32 GB main memory. The CPU we use is Intel(R) Core(TM) i7-5960X CPU @ 3.00GHz (8 cores with 16 threads). The operating system is Ubuntu 16.04.4 LTS 64-bit. The \texttt{C++} compiler is g++ 5.4.0.

\vspace{0.1in}
\noindent\textbf{Retrieval quality.} Recall is a widely-utilized retrieval quality measurement in approximate nearest neighbor search. Consider the top-$K$ result set returned by an algorithm is $A$, and the correct $K$ nearest neighbor set of the query is $B$. The recall is computed as:
\begin{equation*}
Recall(A) = \frac{|A\cap B|}{|B|}.
\end{equation*}
A higher recall represents a better retrieval from the correct nearest neighbor result. Note that $|A|=|B|=K$ in this measurement. Therefore, recall and precision in this context are equivalent.

\vspace{0.1in}
\noindent\textbf{Evaluation measure.} We use Recall vs. Time (\textbf{Queries Per Second (QPS)})) to evaluate the searching performance.
Recall vs. Time reports how many queries the algorithm can process per second at each recall level. Both evaluated method, SL2G and GUITAR have three common parameters: $M$, $k_{\text{construction}}$ and $k_{\text{search}}$, which control the degrees of vertices and the number of search attempts. Note that GUITAR is a searching framework---we use exactly the same proximity graph built with SL2G. To make a fair comparison, we vary these parameters over a fine grid. For each algorithm in each experiment, we will have multiple points scattered on the plane. To plot  curves, we first find out the best recall number, \textit{max-recall}. Then 100 buckets are produced by splitting the range from $0$ to \textit{max-recall} evenly. For each bucket, the best result along the other axis (e.g., the highest queries per second or the lowest percentage of pair-wise computations) is chosen. If there are no data points in the bucket, the bucket will be ignored. In this way, we shall have multiple pairs of data for drawing curves.

\vspace{0.1in}
\noindent\textbf{GUITAR versus SL2G.}
Figure~\ref{fig:exp:guitar-sl2g} reports the comparison between GUITAR and SL2G on Twitch and Amazon datasets. The GUITAR here uses the separation angle based ranking and pruning. The tolerance factor $\alpha = 1.01$---the neighbor vertices, whose separation angles to the reverse direction of the gradient locate within $1\%$ difference of the smallest separation angle, are selected in the probable candidate set.
We can observe a clear gap between GUITAR and SL2G for all settings with different Top-K recalls on both datasets. Our proposed GUITAR searching algorithm outperforms SL2G for all recall levels. For moderate top-1 recall levels, e.g., $80\%$, GUITAR is around $2.7X$ faster than SL2G on both Twitch and Amazon dataset. When we try to retrieve more and more results (top-10, 50, and 100), a more than $2X$ speedup can still be observed, e.g., for the top-100 recall, the speedup of GUITAR is 2.2X on Twitch and 2.6X on Amazon. For high recall levels, e.g., $95\%$, GUITAR consistently shows a $2-4$ times performance improvement over the previous state-of-the-art method, SL2G.

\begin{figure}
\begin{center}
\mbox{
\includegraphics[width=1.9in]{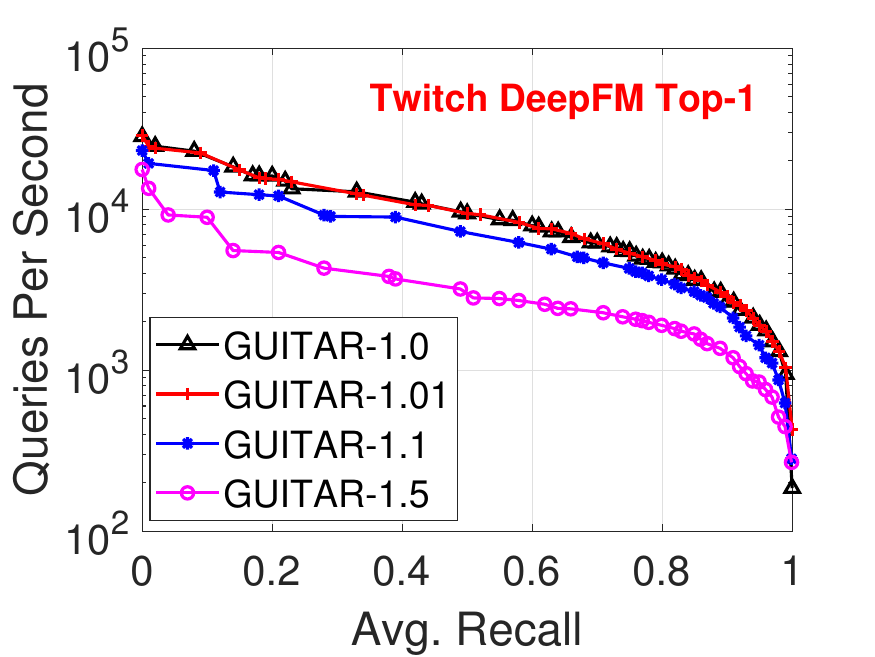}
\hspace{-0.22in}
\includegraphics[width=1.9in]{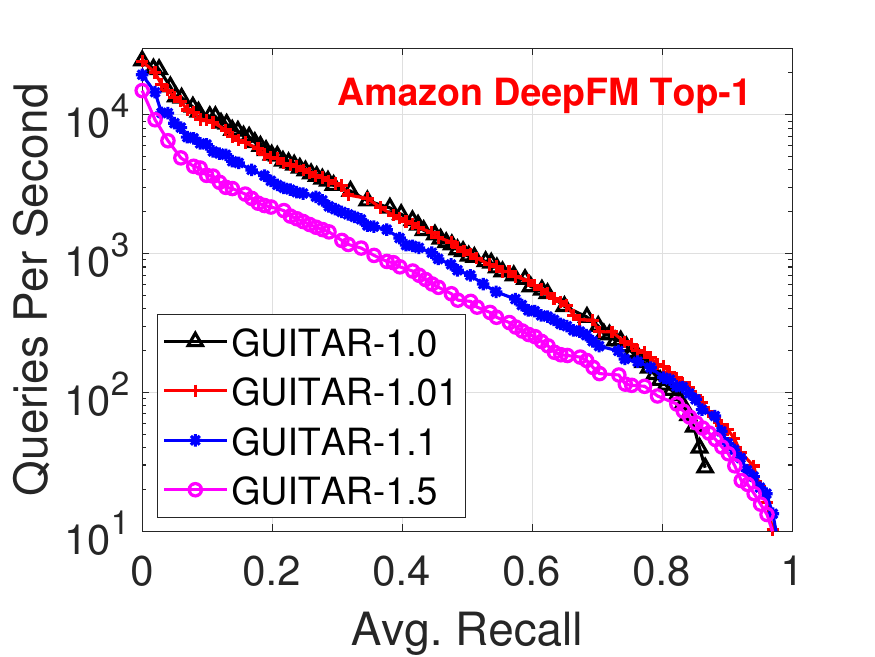}
}
\mbox{
\includegraphics[width=1.9in]{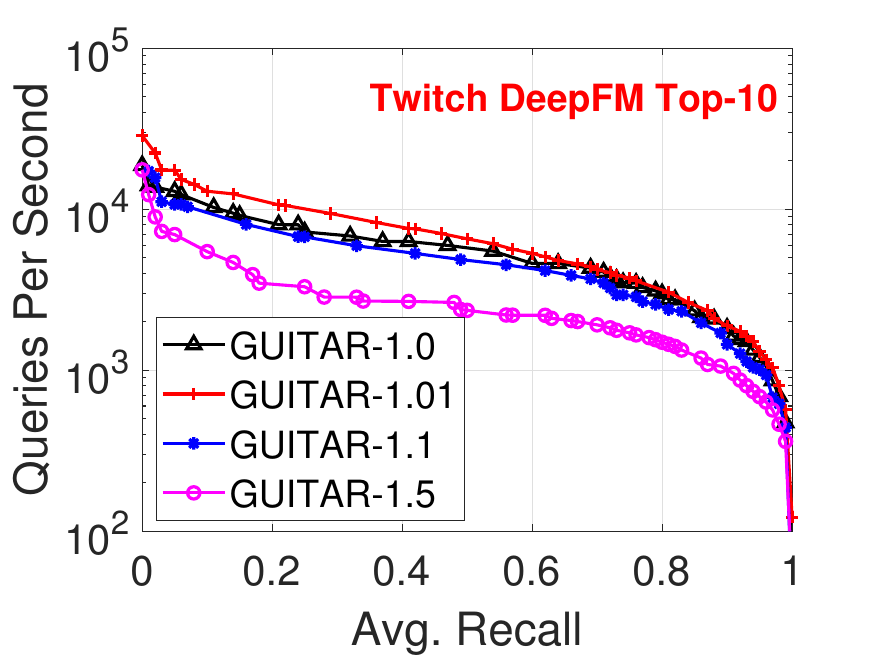}
\hspace{-0.22in}
\includegraphics[width=1.9in]{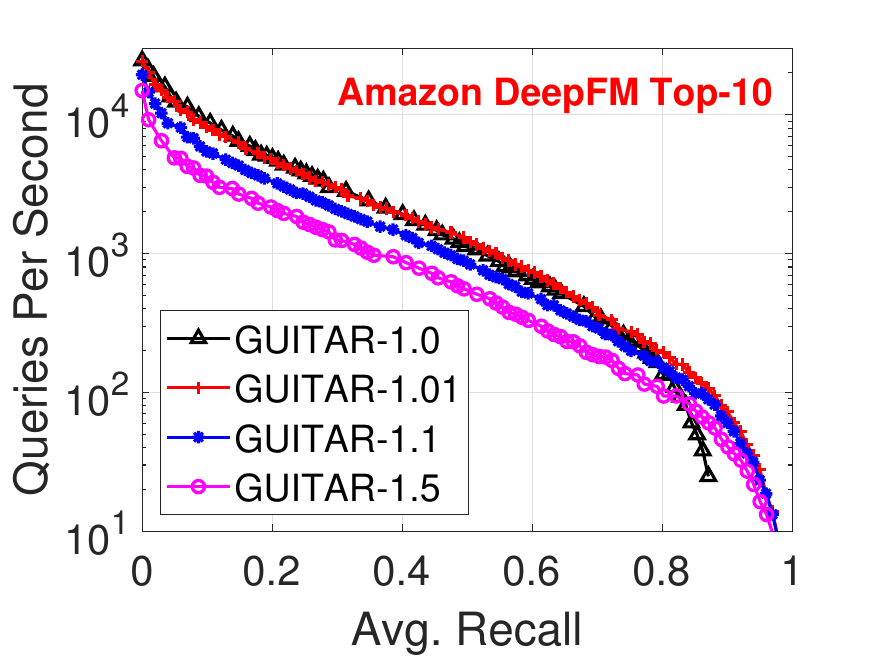}
}
\mbox{
\includegraphics[width=1.9in]{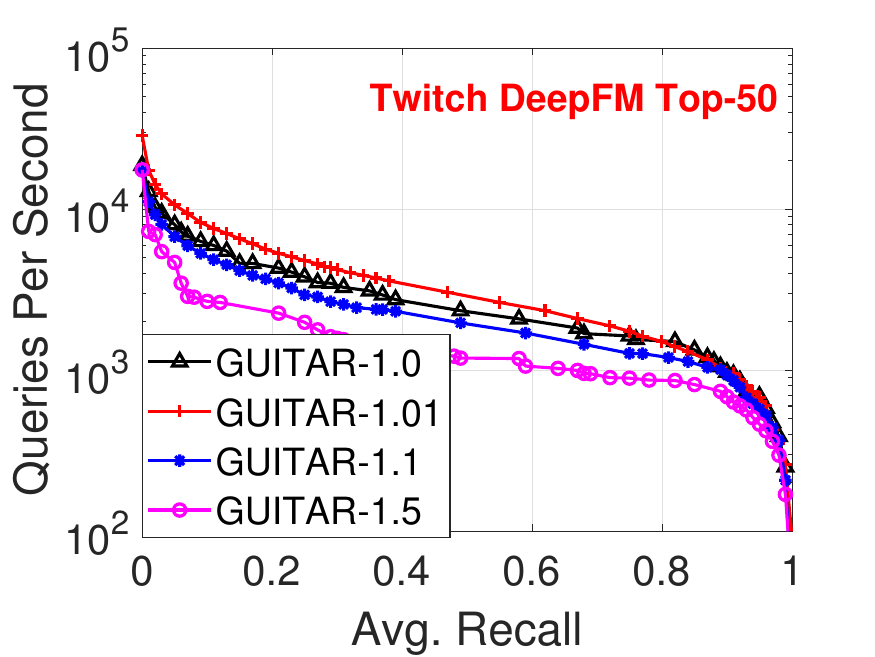}
\hspace{-0.22in}
\includegraphics[width=1.9in]{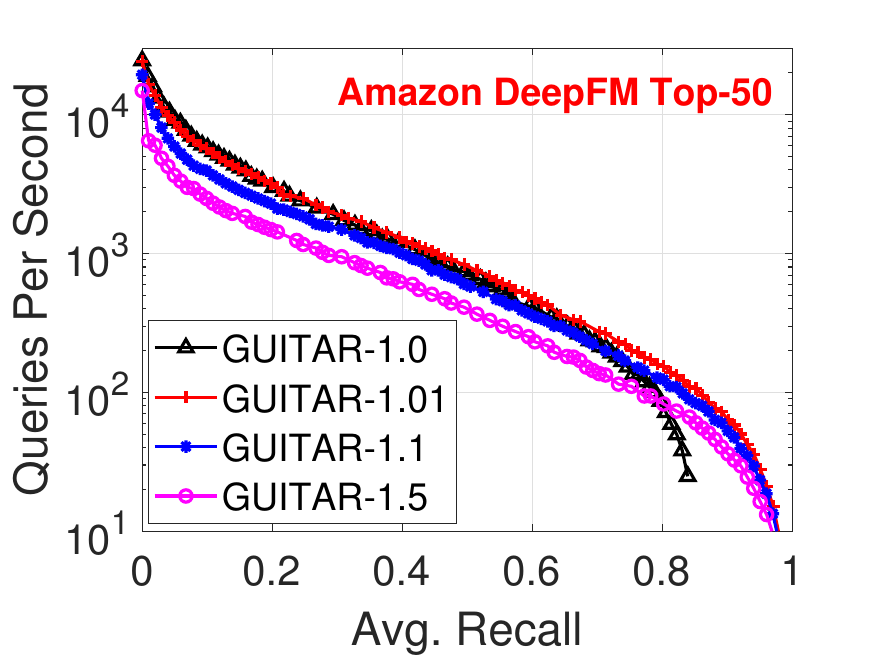}
}
\mbox{
\includegraphics[width=1.9in]{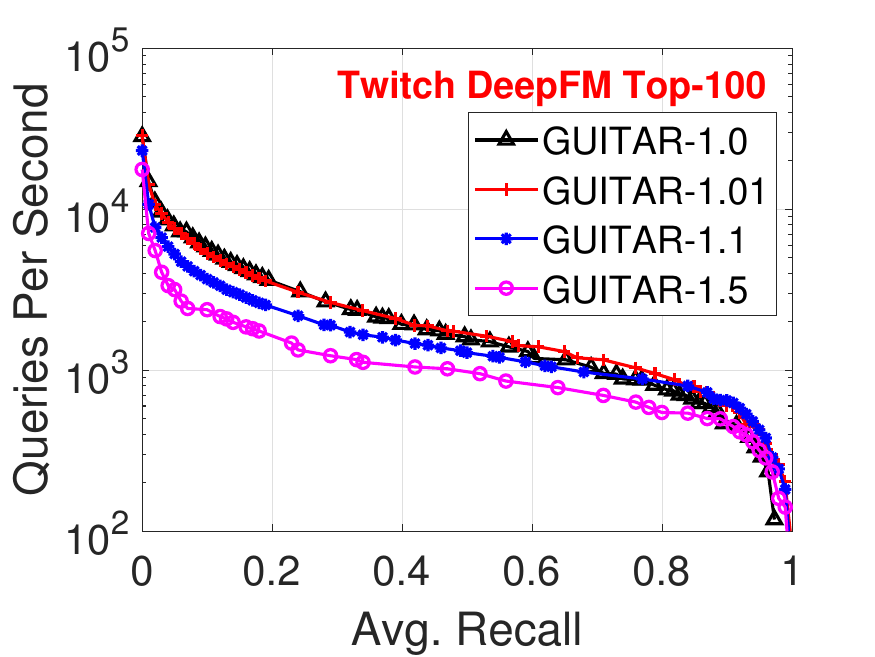}
\hspace{-0.22in}
\includegraphics[width=1.9in]{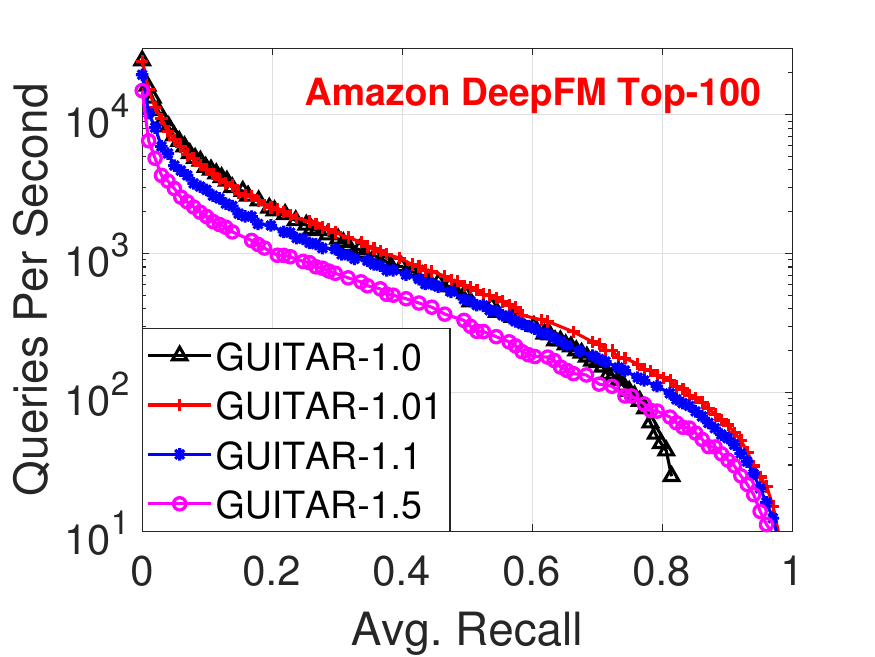}
}

\end{center}

 \caption{Performance comparison when using 1.0, 1.01, 1.1, and 1.5 as the tolerance factor $\alpha$ for the separation angle ranking and pruning strategy. The best results are in the upper right corner.
}\label{fig:ratio}
\end{figure}

\vspace{0.1in}
\noindent\textbf{Tolerance factor $\alpha$.}
Figure~\ref{fig:ratio} depicts the comparison when we vary the tolerance factor $\alpha$ for the separation angle ranking and pruning strategy. We only show the two representative cases (top-1 and 100) here. The top-10 and 50 cases follow the same trend and thus are omitted. When $\alpha$ is large, i.e., $1.5$, we can observe a clear performance drop because a large $\alpha$ includes unnecessary vertices that are not likely to be the top-k results into the probable candidate set---it results in unnecessary neural network evaluations. When $\alpha$ is sufficiently small, e.g., $1.0$, $1.01$, and $1.1$, the performance impact of $\alpha$ is minimal. A strict pruning constraint ($\alpha = 1.0$) has a superior performance for low recall levels in the top-1 case. While we target to get high recall levels or retrieve more top-k results (top-100), $1.0$ over-prunes neighbor vertices, thus it leads to a noticeable performance drop, e.g., the searching performance of $\alpha=1.0$ declines around recall $80\%$ for the Amazon top-100 case. Overall, we suggest using a small $\alpha$, i.e., $1.01$, that reveals a consistent performance.

\newpage

\noindent\textbf{Pruning Benefits versus Overhead.}
We show the computation breakdown of the neural network evaluations and the gradient computation costs in Table~\ref{tbl:nncompute}.

\begin{table}[h]
\centering
\caption{Computation Breakdown on Twitch: R@100 means the top-100 recall; \#NN is the number of neural network evaluations; \#Grad is the number of gradient computations. We use the same proximity graph with $k_{\textit{construction}}=100$ and $M=24$. Numbers with the best performance are highlighted in \textbf{bold}. $*$Total for SL2G is the times of going through the neural network (i.e., $\#$NN) and for GUITAR is $\#$NN+$\#$Grad$\times$2.\vspace{0.1in}}
\begin{tabular}{|c|l|r|r|r|r|} \hline
R@100 & Methods&\multicolumn{1}{c|}{\#NN} &\multicolumn{1}{c|}{\#Grad} & \multicolumn{1}{c|}{Total$^{*}$} & \multicolumn{1}{c|}{QPS}\\ \hline
\multirow{4}{*}{$85\%$} & SL2G & 741.49 &- &741.49 &426.69 \\
& GUITAR-1.0 & 192.03&190.36 &572.75 & 572.76\\
& GUITAR-1.01 & 179.95&132.53 &\textbf{445.01} & \textbf{767.66}\\ & GUITAR-1.1 & 305.55&99.76 &505.07 & 627.37\\
& GUITAR-1.5 & 444.29&89.51 &623.31 & 469.40\\ \hline
\multirow{4}{*}{$90\%$} & SL2G & 983.02 &- &983.02 & 319.53 \\
& GUITAR-1.0 & 235.82&234.24 & 704.30& 465.07\\
& GUITAR-1.01 & 234.60&171.95 &\textbf{578.50} & \textbf{584.28}\\ & GUITAR-1.1 & 365.66&119.53 &604.72 & 543.19\\
& GUITAR-1.5 & 526.04&109.31 &744.66 & 384.28\\ \hline
\multirow{4}{*}{$95\%$} & SL2G & 2125.34 &-  &2125.34 &148.03 \\
& GUITAR-1.0 & 698.69&698.06  &2094.81 &151.82\\
& GUITAR-1.01 & 401.03&291.23 &\textbf{983.49} & \textbf{347.31}\\ & GUITAR-1.1 & 575.99&219.12 &1014.23 & 295.83\\
& GUITAR-1.5 & 804.37&178.76 & 1161.89& 252.77\\ \hline
\end{tabular} \label{tbl:nncompute}
\end{table}

We report the statistics for three top-100 recalls, $85\%$, $90\%$, and $95\%$ on the dataset of Twitch. The compared methods are SL2G and GUITAR with $\alpha=1.0$, $1.01$, $1.1$, and $1.5$. Per query, \#NN denotes the average number of neural network measure evaluations, \#Grad shows the average number of gradient computations, and QPS is the abbreviation of Query Per Second (a higher QPS corresponds to a better performance). Note that the column ``Total'' in the table is the weighted summation of \#NN and \#Grad (Total = \#NN + \#Grad * 2)---the time to compute a gradient is around two times of the time to evaluate one neural network, because gradient computation requires one feed forward and one back propagation.  Although one gradient computation costs two times of a neural network evaluation, GUITAR saves a substantial proportion of neural network evaluation---the column ``Total'' shows the total number of times we go over the neural network (one neural network evaluation counts 1 and one gradient computation counts 2).
As we analyzed in Section~\ref{sec:rank}, the cost of the ranking and pruning in GUITAR is dominated by the number of neural network evaluations---QPS is negatively correlated to ``Total''. Based on this table, we can conclude that the bi-level searching algorithm in GUITAR gains substantially more performance improvement than the overhead we paid to rank and prune the neighbor vertices.

\begin{figure}
\begin{center}
\mbox{\hspace{-0.15in}
\includegraphics[width=1.9in]{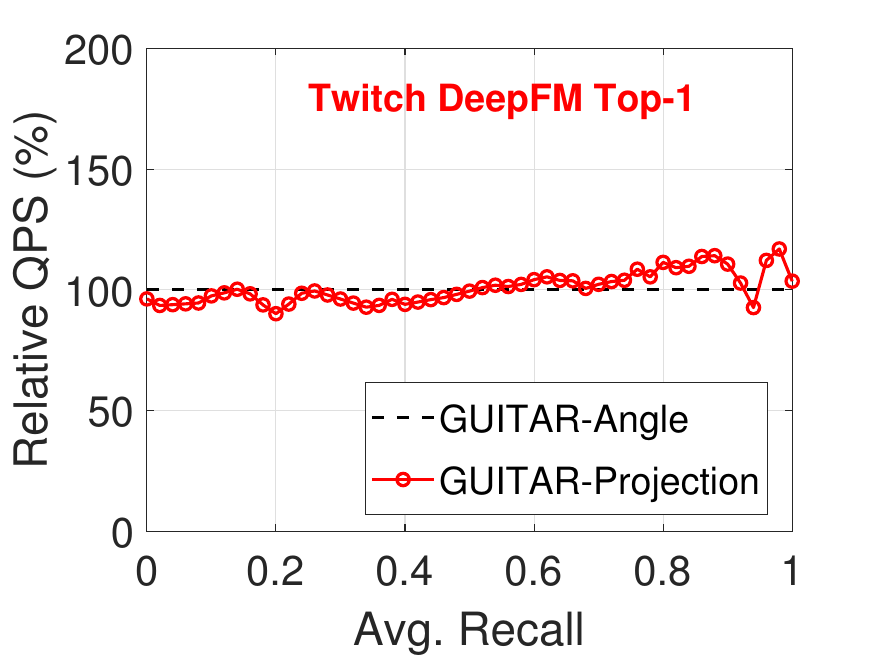}
\hspace{-0.22in}
\includegraphics[width=1.9in]{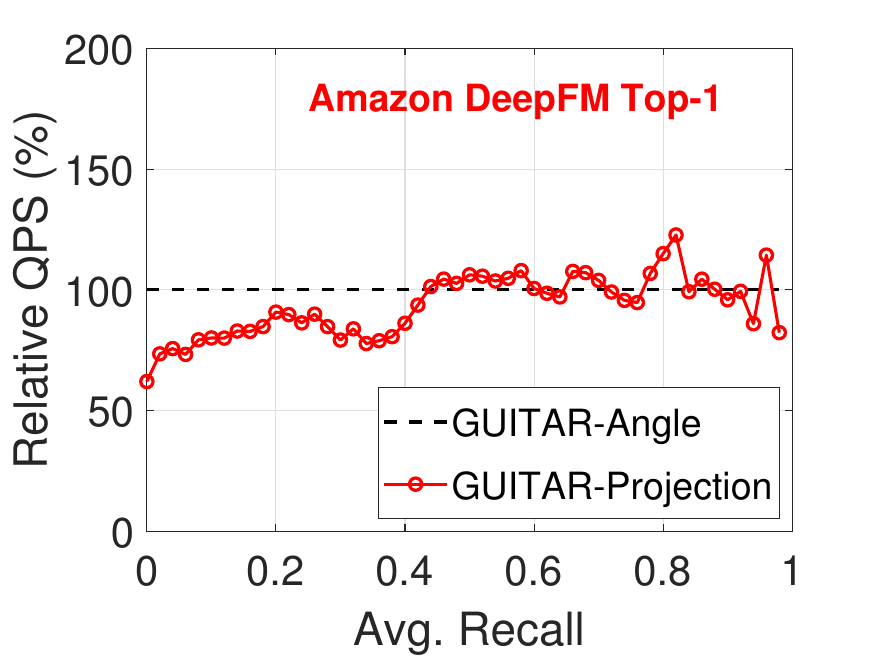}
}
\mbox{\hspace{-0.15in}
\includegraphics[width=1.9in]{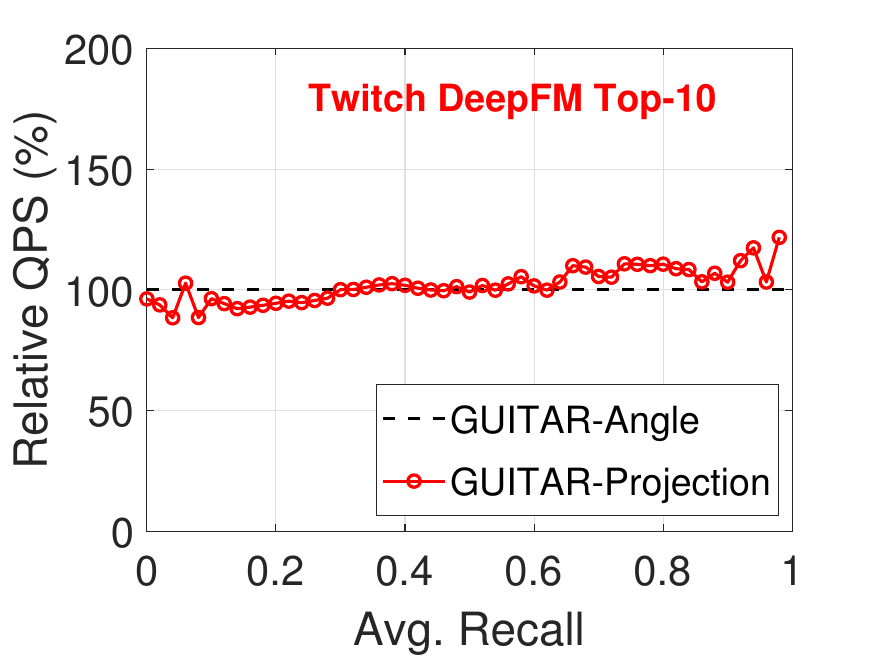}
\hspace{-0.22in}
\includegraphics[width=1.9in]{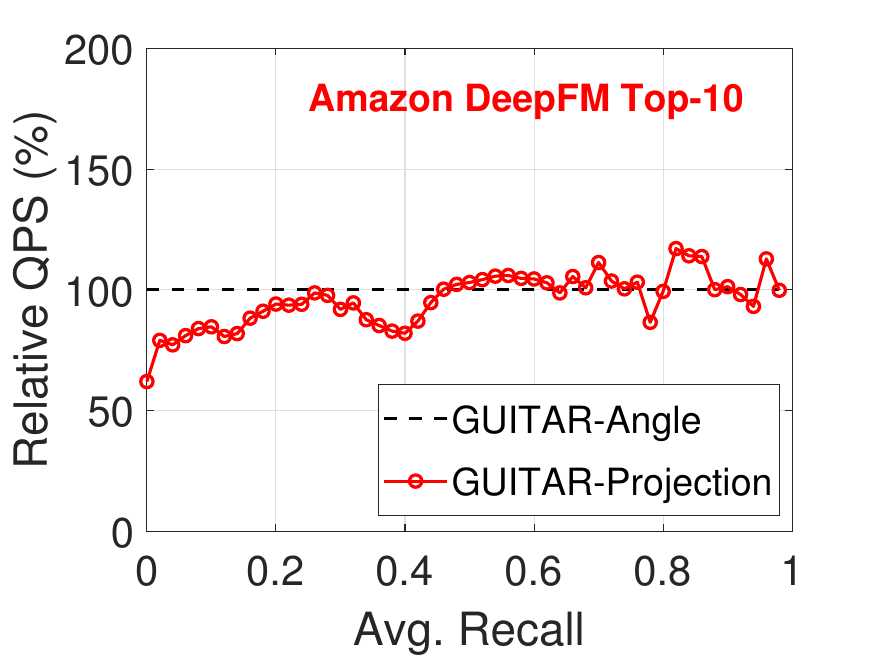}
}
\mbox{\hspace{-0.15in}
\includegraphics[width=1.9in]{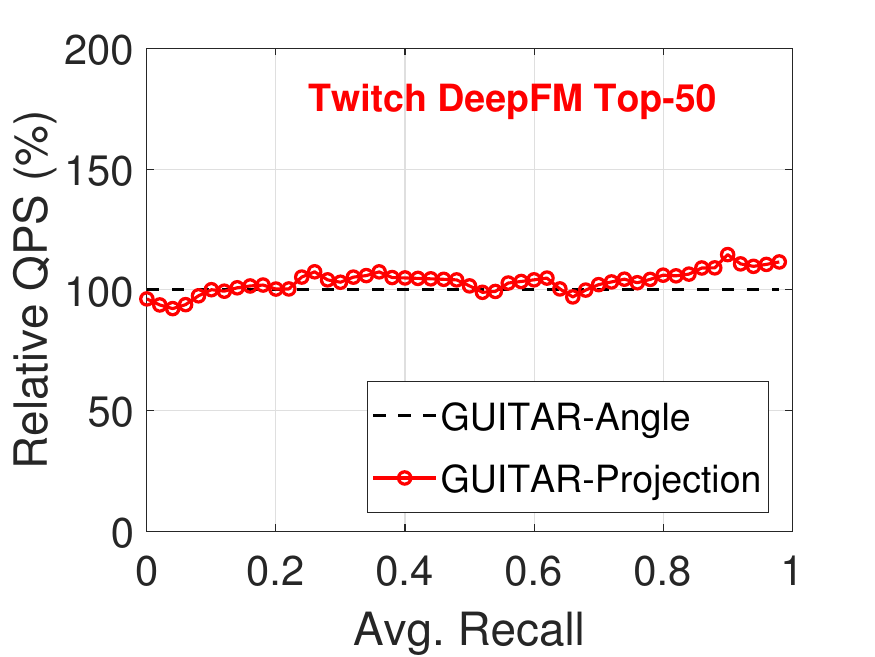}
\hspace{-0.22in}
\includegraphics[width=1.9in]{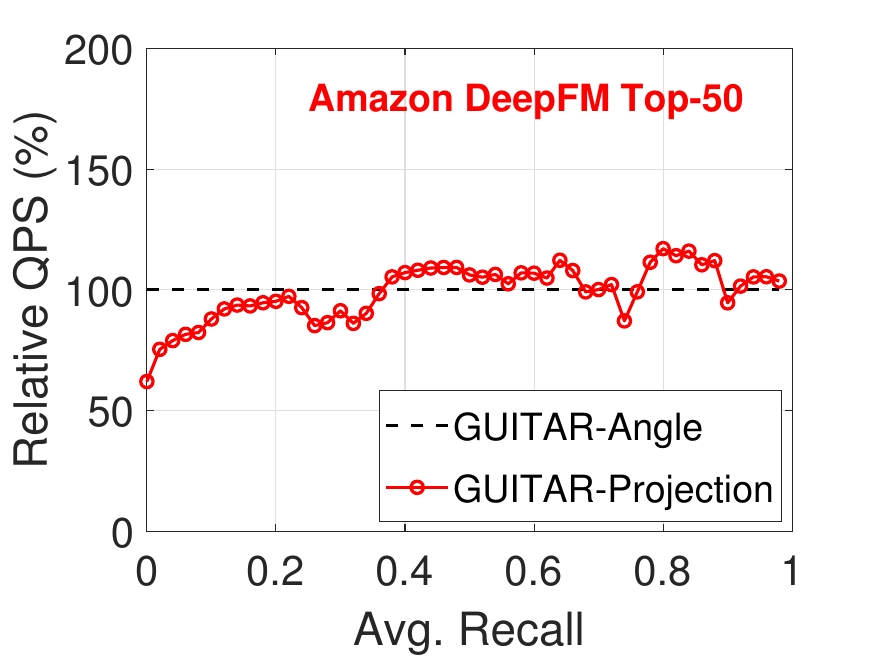}
}
\mbox{\hspace{-0.15in}
\includegraphics[width=1.9in]{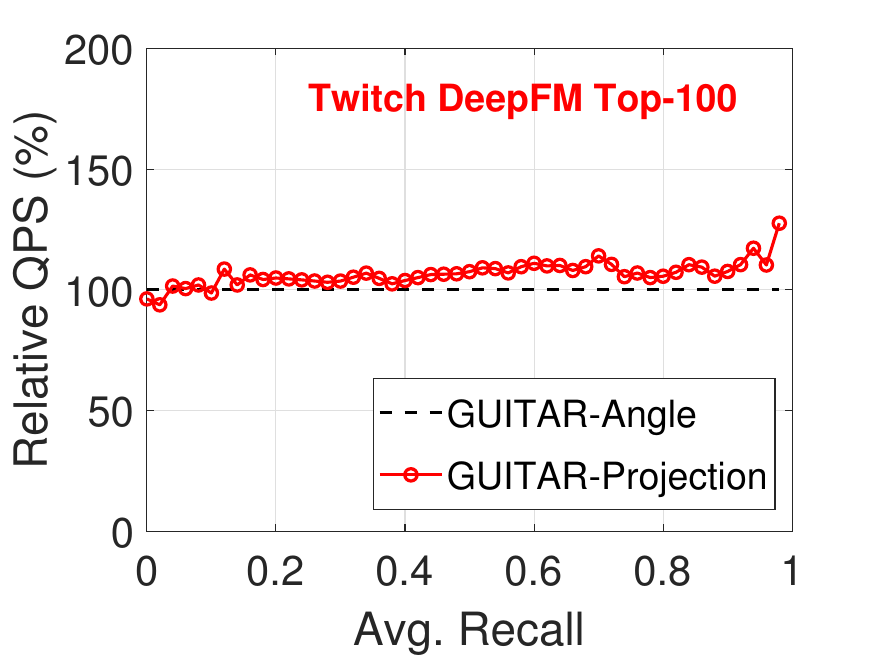}
\hspace{-0.22in}
\includegraphics[width=1.9in]{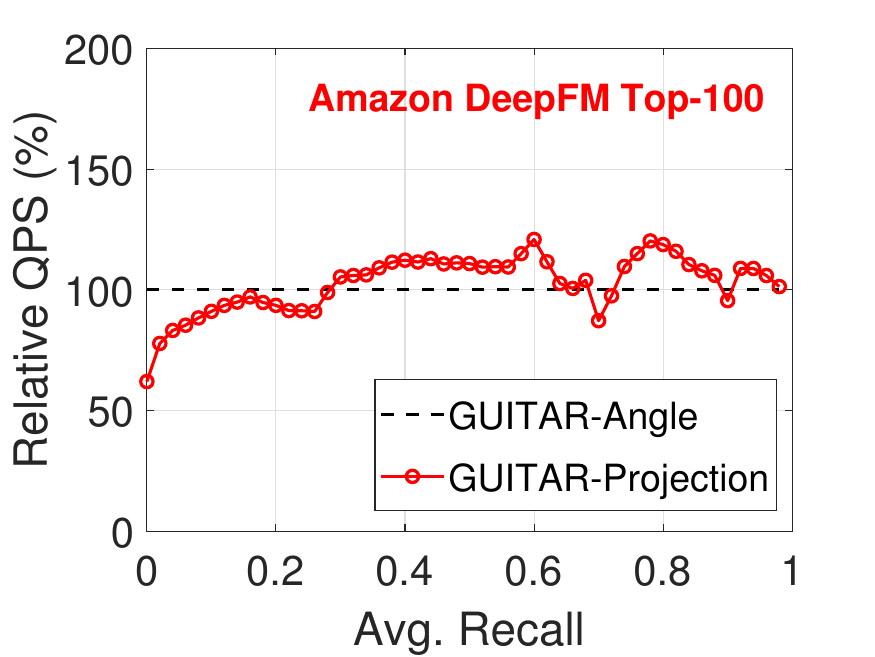}
}
\end{center}


 \caption{Performance of projection based ranking and pruning strategy: relative QPS to the separation angle based one. The best results are in the upper right corner.
}\label{fig:proj} 

\end{figure}

\vspace{0.1in}
\noindent\textbf{Separation angle versus projection.}\label{sec:exp:project}
We compare the alternative projection based ranking and pruning strategy with the separation angle based one in Figure~\ref{fig:proj}. The plots show the relative QPS to the baseline separation angle based ranking and pruning strategy with $\alpha=1.01$ (which consistently shows a good performance in our experiments). We enumerate the tolerance $\alpha$ to obtain the best QPS for the projection based method, where the best $\alpha$ we found for Twitch is $2.0$ and $10.0$ for Amazon.
For most recall levels, the differences between both methods are within $10\%$ for Twitch and $20\%$ for Amazon. The projection based strategy gains some marginal improvement over the separation angle based one for $>50\%$ recalls. However, the relative QPS fluctuates in high recall ranges. In general, both strategies have similar performance.

\begin{figure}[h!]
\begin{center}
\mbox{\hspace{-0.05in}
    \includegraphics[width=1.9in]{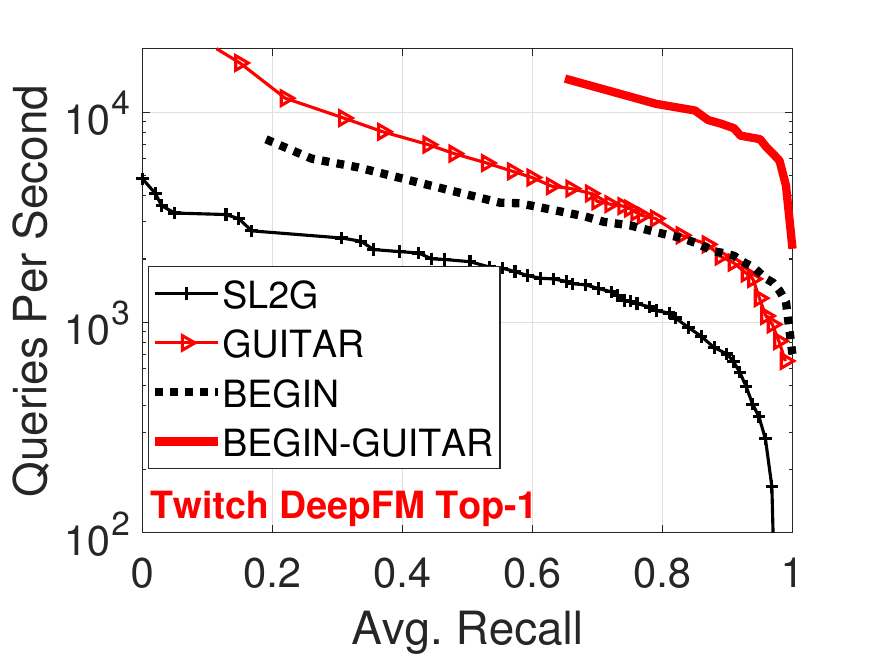}
    \hspace{-0.15in}
    \includegraphics[width=1.9in]{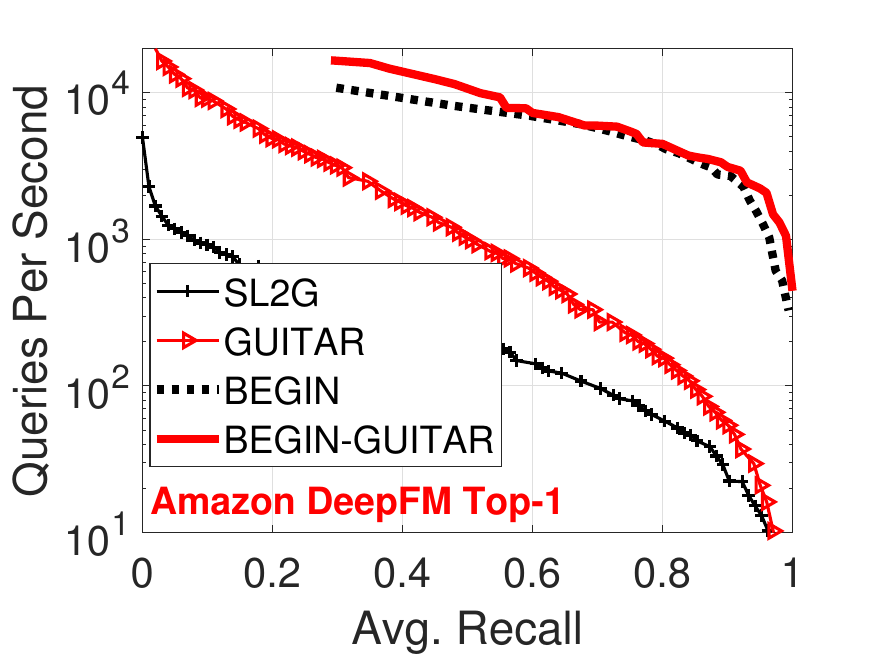}
    }
\mbox{\hspace{-0.05in}
    \includegraphics[width=1.9in]{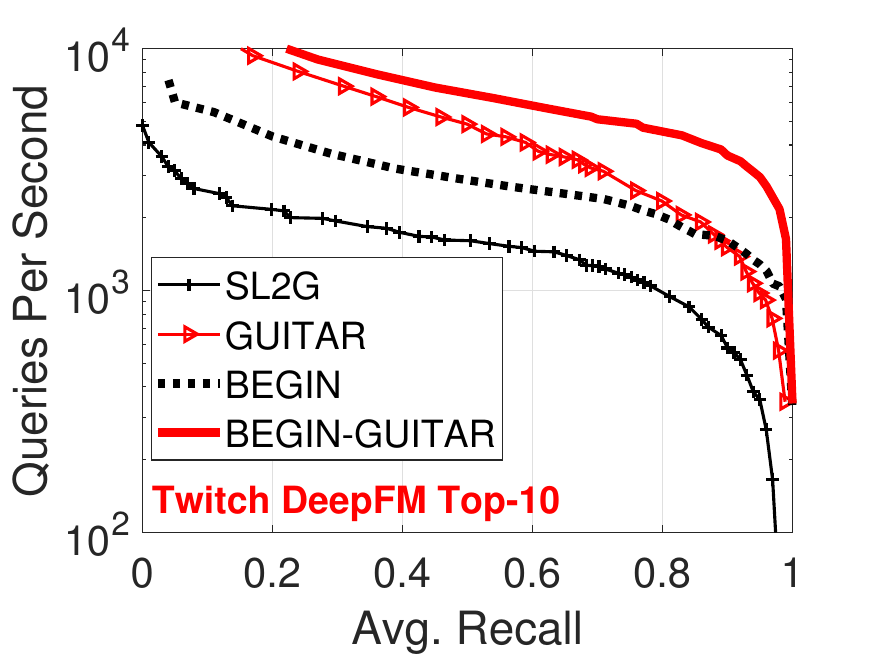}
    \hspace{-0.15in}
    \includegraphics[width=1.9in]{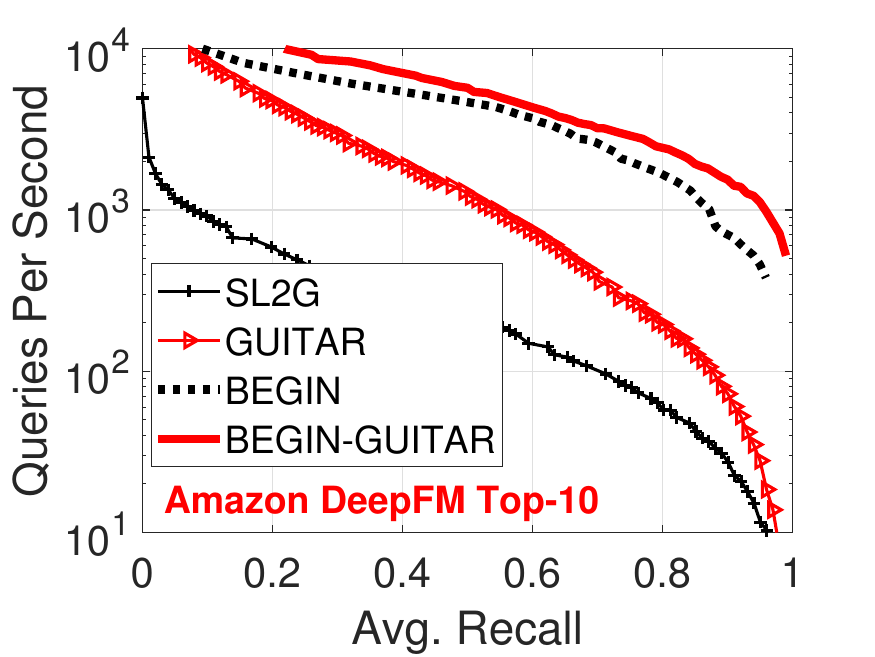}
    }
\mbox{\hspace{-0.05in}
    \includegraphics[width=1.9in]{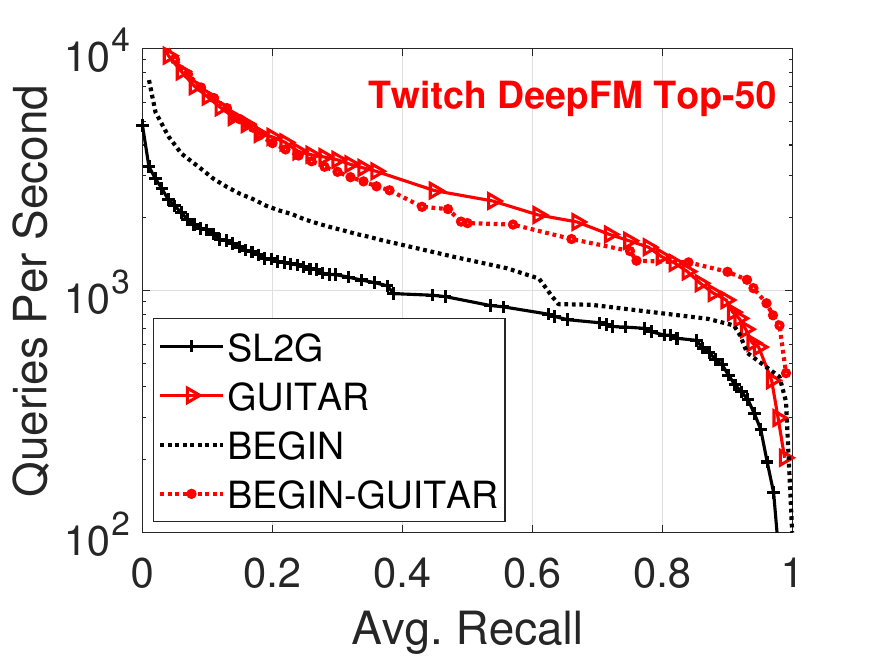}
    \hspace{-0.15in}
    \includegraphics[width=1.9in]{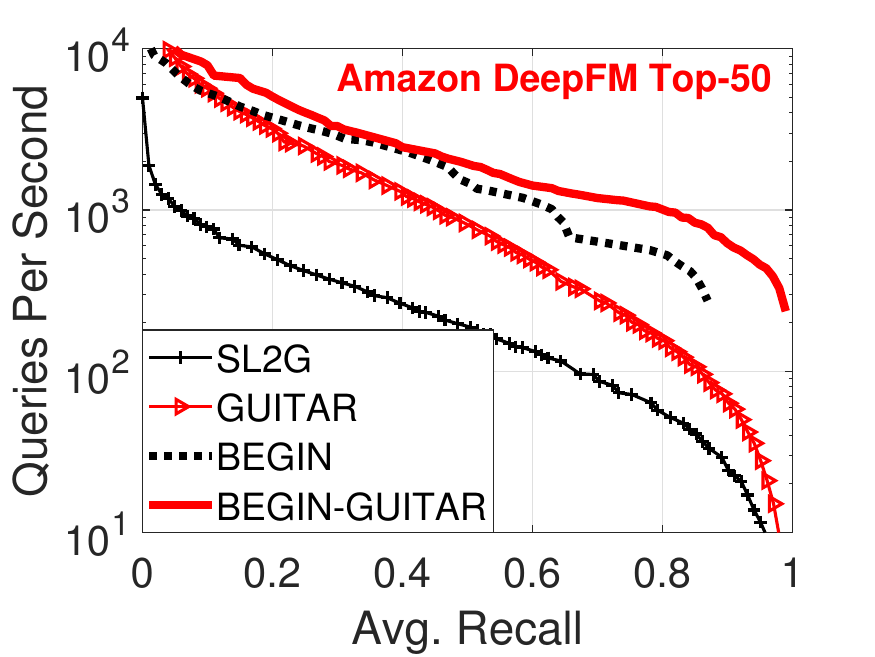}
    }
\mbox{\hspace{-0.05in}
    \includegraphics[width=1.9in]{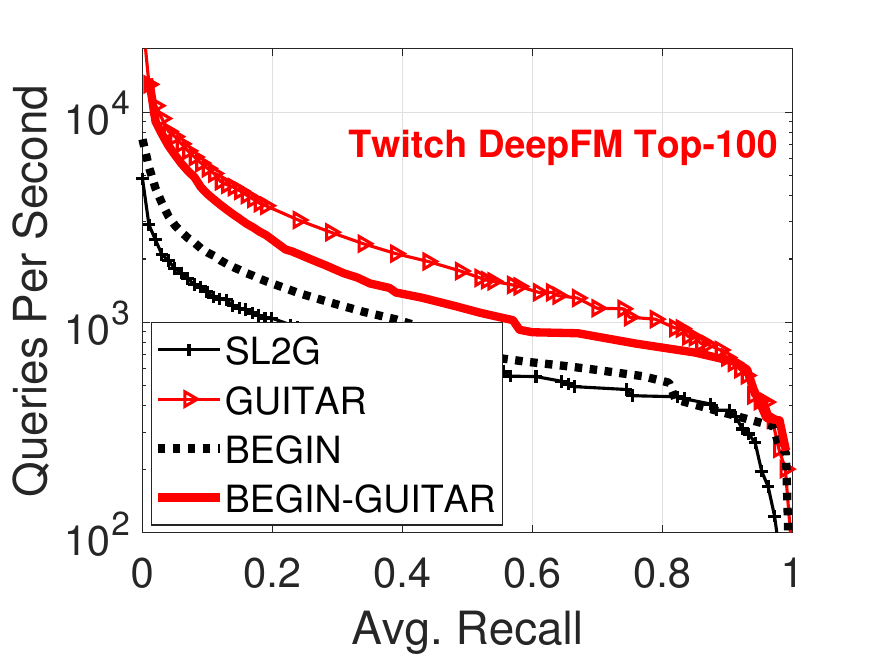}
    \hspace{-0.15in}
    \includegraphics[width=1.9in]{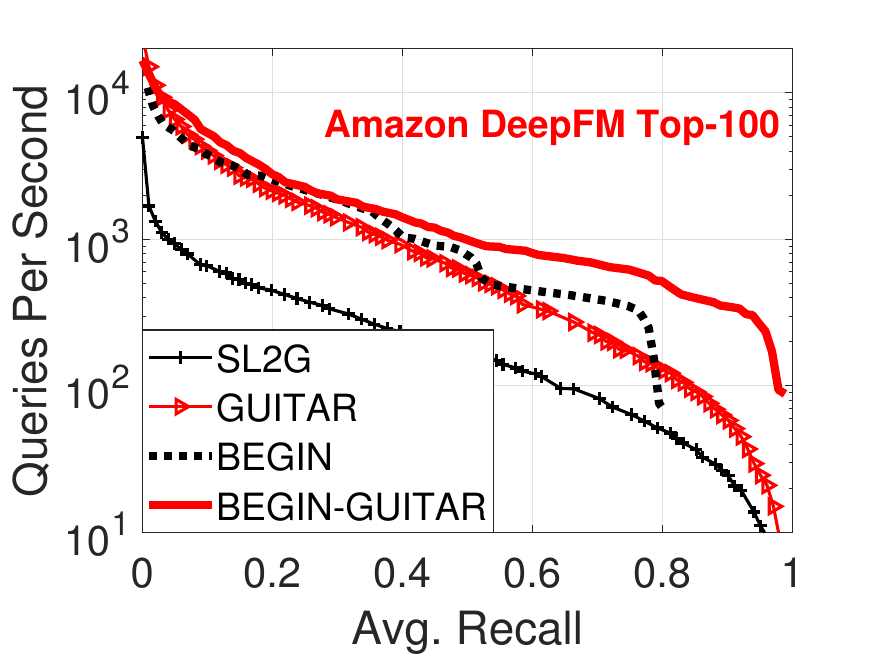}
    }
\end{center}

\caption{Performance comparison for SL2G, GUITAR, BEGIN, and BEGIN-GUITAR. The best results are in the upper right corner.}
\label{fig:exp:guitar-sl2g-begin}
\end{figure}

\vspace{0.1in}
\noindent\textbf{Adaption to BEGIN.}
BEGIN~\citep{tan2021fast} constructs a bipartite graph to consider the neural network $f$ in the index building stage---it takes more offline time to build the index than SL2G, but yields a better graph index for searching. Our proposed GUITAR in this paper is a searching method on the graph. Therefore, it can also be adapted to the bipartite graph index generated by BEGIN: we can straightforwardly use the GUITAR gradient pruning technique on the bipartite graph searching. Figure~\ref{fig:exp:guitar-sl2g-begin} reports the performance comparison of GUITAR on the BEGIN bipartite graph index. GUITAR-BEGIN shows the performance when we use the proposed GUITAR pruning over the bipartite graph generated by BEGIN. The GUITAR performance on Twitch Top-1, Top-100, and Amazon Top-100 can be comparable or better than the bipartite graph from BEGIN. Note that building the BEGIN graph requires many time-consuming neural network evaluations while the vanilla GUITAR works on the SL2G index which can be quickly built with simple Euclidean distance computations. Furthermore, we can use the GUITAR searching algorithm over the graph generated by BEGIN for more competitive query performance (Twitch Top-1, Amazon Top-1 and Top-100): GUITAR-BEGIN consistently outperforms over BEGIN.

\vspace{.15in}
\section{Conclusions}

The ultimate goal of the fast neural ranking is to merge the first-stage item retrieval and the following re-ranking stage with a complicated neural network into one step. Using a more complex measure may have better effectiveness and lower QPS (higher throughput and lower latency). We believe the impact of our work is to bridge the gap of this trade-off.

In this paper, we introduce a novel graph searching framework to accelerate the searching in the fast neural ranking problem. The proposed graph searching algorithm is bi-level: we first construct a probable candidate set; then we only evaluate the neural network measure over the probable candidate set instead of evaluating the neural network over all neighbors.
We propose a gradient-based algorithm that approximates the rank of the neural network matching score to construct the probable candidate set and present an angle-based heuristic procedure to adaptively identify the proper size of the probable candidate set. We extensively evaluate our proposed systems on real data and real neural network measures. The experimental results on public datasets show that GUITAR consistently outperforms the baseline methods.

\balance
\bibliographystyle{plainnat}
\bibliography{refs_scholar}

\end{document}